\documentclass[aps,prd,showpacs,amssymb,superscriptaddress, notitlepage,floats,floatfix,nofootinbib, twocolumn]{revtex4-1}

\usepackage{graphicx}
\usepackage{dcolumn}
\usepackage{bm}
\usepackage{amsmath}
\usepackage{color}
\usepackage{hyperref}
\usepackage{journals}

\begin{document}

\title{Bayesian Metric Reconstruction with Gravitational Wave Observations} 

\author{Sebastian H. V\"olkel}
\email{sebastian.voelkel@sissa.it}
\author{Enrico Barausse}
\email{barausse@sissa.it}

\affiliation{SISSA, Via Bonomea 265, 34136 Trieste, Italy and INFN Sezione di Trieste}
\affiliation{IFPU - Institute for Fundamental Physics of the Universe, Via Beirut 2, 34014 Trieste, Italy}

\date{\today}

\begin{abstract}
Theories of gravity extending or modifying general relativity typically allow for black hole solutions different from the Schwarzschild/Kerr geometries. Electromagnetic observations have been  used to place constraints on parametrized deviations from the Schwarzschild/Kerr metrics, in an effort to gain insight on the underlying gravitational theory. In this work, we show that observations of the gravitational quasi-normal modes by existing and future interferometers can be used to bound the same parametrized black hole metrics that are constrained by electromagnetic observations (e.g. by the Event Horizon Telescope). We argue that our technique is most sensitive to changes in the background black hole metric near the circular photon orbit, and that it is robust against the changes that a gravitational theory differing from general relativity necessarily introduces in the equations for the gravitational perturbations. We demonstrate our approach by reconstructing the background metric from a set of simulated observations using a Bayesian approach.
\end{abstract}

\maketitle

\section{Introduction}\label{introduction}
The increasingly precise observations of the gravitational wave signals emitted by merging compact objects 
provide unprecedented opportunities to test general relativity (GR) and the nature of black holes and 
neutron stars~\cite{PhysRevLett.116.061102,PhysRevLett.116.221101,PhysRevLett.116.241103,PhysRevLett.118.221101,Abbott_2017,PhysRevLett.119.141101,PhysRevLett.119.161101,Abbott_2020,LIGOScientific:2020stg}. 
Among the predictions of GR (and also other gravitational theories) is the existence of quasi-normal modes (QNMs), 
which describe the characteristic spacetime oscillations of perturbed compact objects. 
These (damped) oscillations can be observed after the violent merger of two compact objects.
In this phase, the so-called ringdown, the final remnant forming from the coalescence 
settles into an equilibrium stationary configuration by radiating in  QNMs. 
If the final object is a black hole, the no-hair theorem of GR~\cite{Carter:1971zc,Robinson:1975bv} states that 
it must be described by the Kerr geometry~\cite{Kerr:1963ud}, which is 
 fully characterized by the mass $M$ and spin $J$, with the 
latter satisfying the ``Kerr bound'' $|J|\leq M^2$ (in the units $G=c=1$ that we utilize throughout this paper) 
to avoid the presence  of naked singularities.

While the no-hair theorem holds in GR, 
gravitational theories modifying and/or extending it generally yield different black hole spacetimes~\cite{berti_review},
and also different equations for the gravitational perturbations over the background geometry~\cite{barausse_sotiriou,berti_review}.
One way to test GR is therefore to verify that the observed QNMs from
the remnant of a binary black hole merger match those of a Kerr black hole. 
This is commonly referred to as  ``black hole spectroscopy''~\cite{Berti_2009}, and may become feasible 
with future ground or space-based gravitational wave detectors~\cite{Berti:2016lat,Shi:2019hqa}, or even with current data~\cite{Ghosh:2017gfp,Brito_2018,Carullo:2018sfu,Isi:2019aib,PhysRevLett.116.221101}.

From a practical point of view, there are different approaches to the problem. 
The first one is ``top-down'', and consists  of choosing a specific theory of gravity, finding black hole solutions, 
deriving the gravitational perturbation equations, and computing the QNM spectrum. 
While this allows one to make precise predictions based on a specific theory, it requires several non-trivial steps, and generally only provides insight  on
the very theory under investigation. 
A second ``bottom-up'' approach is supposed to be as theory agnostic as possible, and assumes a parametrized working ansatz for the black hole 
background metric  as a starting point. 
One can then compute observables
that depend on the background metric alone, e.g. motion of small bodies 
with weak internal gravity, which  are relevant  for  instance for
extreme mass-ratio  inspirals~\cite{Glampedakis:2005hs} and which follow geodesics of the background metric
in gravitational theories that satisfy the weak equilavence principle~\cite{Will:2018bme,Barausse:2016eii}.

Note however that the computation of the gravitational QNM spectrum cannot be performed easily in this second approach, because of the lack of field equations. 
A possibility would be to parametrize the field equations as well. For instance, Refs.~\cite{PhysRevD.100.044040,glampedakis} consider
scalar tensor theories, which are the simplest extension of GR. These theories include
 an extra scalar graviton polarization, which Refs.~\cite{PhysRevD.100.044040,glampedakis} couple (via free parameters) to the tensor gravitons of GR. 
 Refs.~\cite{PhysRevD.100.044040,glampedakis} then study the QNMs of this coupled scalar and tensor system over generic spherical and axisymmetric backgrounds, 
 in the eikonal limit (see also Refs.~\cite{Carson:2020iik,Carson:2020ter} for similar attempts). While this formalism is very general, in this work we will follow a simpler approach. In more detail,
we will look at the axial sector of the gravitational QNMs of a spherically  symmetric and static parametrized black hole metric. The reason
for focusing on the axial sector is that at linear order, the scalar perturbations cannot mix with
the axial gravitational perturbations, because of parity. As a result, the equation for the
axial gravitational QNMs in generic scalar tensor theories is expected to be the same as the Regge-Wheeler equation of GR \cite{PhysRev.108.1063} (at  least in the eikonal limit) although on a background differing from Schwarzschild and with a modified potential.
Note that it may be possible to extend this approach to the polar sector too, by resorting to 
an effective field theory treatment such as that of Ref.~\cite{Franciolini:2018uyq}. An analysis of scalar perturbations on modified black-hole metrics obtained within effective field theories can be found in Ref.~\cite{Cano:2020cao}.

The  main goal of this paper is then to address the question of 
how many QNM observations (and with what precision) 
one would need to reconstruct a parametrized black hole metric. 
This problem was qualitatively tackled in Ref.~\cite{paper8} by using
a scalar QNM toy model with the metric proposed by Rezzolla and Zhidenko (RZ)~\cite{PhysRevD.90.084009} in the spherical and static limit (and 
later generalized to the axisymmetric case by Ref.~\cite{Konoplya:2016jvv}). 
By studying how much RZ QNMs differ from Schwarzschild QNMs when multiple RZ parameters are non-zero, it was possible to investigate a subset of the RZ parameter space in terms of a \textit{direct problem}. 
From the reported results one should expect that the general \textit{inverse problem} is non-trivial, because certain RZ parameter combinations could lead to very similar QNMs, even when they are known with high accuracy.
A different and more general way to approach the inverse QNM problem of different types of non-rotating compact objects has been reported in \cite{paper2,paper5,paper6}.
These works focus on reconstructing the perturbation potential directly from the QNM spectrum by inverting generalized Bohr-Somemrfeld rules, but without direct access to the underlying metric.
\par
Here, we improve on the work presented in Ref.~\cite{paper8} by computing axial gravitational QNMs with a 
 higher order WKB method~\cite{PhysRevD.68.024018}, for the spherical and static parametrized RZ metric.
The RZ metric has proven to be a useful and economic approximation to exact black hole spacetimes in alternative theories of 
gravity \cite{konoplya2020general}, and has been used in different type of applications, e.g., for 
black hole shadows \cite{Younsi:2016azx,Mizuno:2018lxz} and for gravitational wave and X-ray tests of the Kerr spacetime \cite{Cardenas_Avendano_2020}.
Focusing on the fundamental $l=2$ and $l=3$ modes and their overtones, which are expected to dominate the ringdown of black holes 
resulting from a binary mergers,
we construct a Bayesian pipeline allowing for estimating the
parameters of the RZ metric, given a set of QNM observations 
from existing or future gravitational wave detectors.

This work is structured as follows. In Sec.~\ref{methods} we outline all our methods and explain the general framework. 
This setup is applied to different scenarios in Sec.~\ref{applications}. We discuss our findings in Sec.~\ref{discussion}, 
before we present our conclusions in Sec.~\ref{conclusions}. 
\section{Methods}\label{methods}
In this section we introduce the building blocks that define the framework of this work. 
We start with an overview of the RZ metric in Sec.~\ref{RZ-metric} and discuss the equations for its perturbations in Sec.~\ref{perturbation equations}.
The computation of QNMs is described in Sec.~\ref{QNM}.
The different combinations of RZ parameters and the subsets of the QNM spectrum that we consider in this work are introduced in Sec.~\ref{def_spectra}.
A discussion of the range of validity of the RZ parameter space is presented in Sec.~\ref{param_space}.
In Sec.~\ref{noise} we discuss the precision of QNM measurements that can be expected with various
experimental setups, and use that information in the Markov chain Monte Carlo (MCMC) framework that is introduced in Sec.~\ref{pymc3}.
\subsection{The RZ Metric}\label{RZ-metric} 
The RZ parametrized metric was introduced to model spherically symmetric black holes beyond GR, in  a theory agnostic way. 
We summarize its most important properties in the following, but refer to the original publication \cite{PhysRevD.90.084009} for full details.
The RZ metric is given by
\begin{align}
	\text{d}s^2 = -N^2(r) \text{d}t^2 + \frac{B^2(r)}{N^2(r)} \text{d}r^2 + r^2 \text{d} \Omega^2,
\end{align}
with $\text{d}\Omega^2 = \text{d}\theta^2 + \sin^2 \theta \text{d}\phi^2$ and two functions $N(r)$ and $B(r)$, which describe the details of the spacetime.
For further convenience, let us remap  the location of the event horizon $r_0$ into the dimensionless coordinate 
\begin{align}
	x \equiv 1-\frac{r_0}{r},
\end{align}
which ranges from $x=0$ at the event horizon to $x=1$ at spatial infinity. 
Another function $A$ is introduced via
\begin{align}
	N^2 = x A(x),
\end{align}
with $A(x)>0$ for $0 \leq x \leq 1$. 
The two functions $A(x)$ and $B(x)$ are given by
\begin{align}
	A(x) &= 1 - \varepsilon(1-x) + (a_0 -\varepsilon)(1-x)^2 + \tilde{A}(x)(1-x)^3,
	\\
	B(x) &= 1+ b_0(1-x) + \tilde{B}(x)(1-x)^2,
\end{align} 
where
$\tilde{A}(x)$ and $\tilde{B}(x)$ describe
deviations from the Schwarzschild limit.
They are introduced as continued fraction expansion 
\begin{align}
\tilde{A}(x) &= \frac{a_1}{1+\frac{a_2x}{1+\frac{a_3x}{1+\dots}}},
\\
\tilde{B}(x) &= \frac{b_1}{1+\frac{b_2x}{1+\frac{b_3x}{1+\dots}}}.
\end{align}

In the original work~\cite{PhysRevD.90.084009}, it was further shown how knowledge from 
solar system tests of the parametrized post-Newtonian (PPN) metric~\cite{Will:2018bme} 
 constrains the parameters $a_0$ and $b_0$ to very small values.
Indeed, solar system tests imply
\begin{align}
        \varepsilon &= - \left(1-\frac{2M}{r_0} \right),\label{varepsilon}
        \\
        a_0&=\frac{(\beta-\gamma)(1+\varepsilon)^2}{2},
        \\
        b_0&= \frac{(\gamma-1)(1+\varepsilon)}{2},
\end{align}
where the PPN parameters $\beta$ and $\gamma$ are constrained to be of order $ \sim 10^{-4}$~\cite{Will:2018bme}. Therefore, in
particular, one has $|a_0|, |b_0| \sim 10^{-4}$.

Nevertheless, we stress that there is no reason to expect PPN bounds to hold for black hole spacetimes
in theories of gravity that modify or extend GR, since Birkhoff's theorem generally does not hold in these
theories. Examples of theories that reproduce the 1PN metric of GR around stars, but which deviate
from GR at 1PN order in black hole spacetimes include scalar tensor theories. The latter
can present screening mechanisms (e.g. chameleon~\cite{chameleon}, K-mouflage~\cite{kmouflage}, symmetron~\cite{symmetron}, etc.) protecting
 local physics from unwanted scalar effects around stars (therefore passing solar system tests),
while still allowing for the existence of scalar charges and 1PN scalar effects in vacuum (see
e.g. \cite{silva,herdeiro,doneva,dima} for such scalarized black holes).
For this reason, in most of this paper we will {\it not} impose PPN bounds on the RZ metric. However, to
allow for comparison with earlier works \cite{paper8,konoplya2020general,Cardenas_Avendano_2020}, we also present some results for
the case in which $a_0=b_0=0$, corresponding to a RZ metric
matching the Schwarzschild one at 1PN order.

\subsection{Perturbation Equations}\label{perturbation equations}

In general, the equations governing the evolution of
linear gravitational perturbations over a black hole  background
depend on the gravitational theory under consideration. In GR,
the gravitational  field only has two  (tensor) polarizations, whose
properties and spectrum are encoded in the Regge-Wheeler~\cite{PhysRev.108.1063} and Zerilli~\cite{PhysRevD.2.2141} equations
(respectively for odd and even metric perturbations on Schwarzschild)
and in the Teukolsky equation~\cite{teuk} (for Kerr perturbations).
In theories extending GR (see e.g. Ref~\cite{berti_review} for a review), not only
can the background black hole spacetime differ from Schwarzschild/Kerr 
(as modeled in Sec.~\ref{RZ-metric}), but even if the background
is the same as in GR (as may happen in specific theories~\cite{psaltis}),
new polarizations will generally be present and will alter the form 
of the pertubation equations~\cite{barausse_sotiriou}.

Some insight on the form of the perturbation equations when one moves beyond GR
can nevertheless be gained by noting that additional modes (beyond the 
spin-2 tensor gravitons of GR)  will typically be coupled weakly to gravitational wave interferometers
if the gravitational theories under scrutiny obeys experimental bounds on the equivalence principle
(c.f. e.g.~\cite{DS,env_effects,STcollapse}). One may therefore safely focus on
the tensor polarizations, whose coupling to detectors is strongest.

In principle, non-tensor polarizations may couple with the tensor degrees of freedom, e.g. appearing as sources
for the equations governing the latter, but this is not a fundamental obstacle to
computing QNMs (see e.g. Refs.~\cite{glampedakis,mcmanus}). Note also that odd parity perturbations will generally be unaffected
by these couplings, at least in scalar-tensor theories respecting parity (e.g.
Fierz-Jordan-Brans-Dicke-like theories; dilatonic Gauss-Bonnet; Horndeski and beyond Horndeski theories; degenerate
higher order  scalar tensor theories, khronometric theory/Ho\v rava gravity, etc).
This is because scalar perturbations have even parity, and therefore cannot mix with
the odd parity sector of the tensor perturbations at linear order.\footnote{Odd tensor modes can in principle mix with pseudoscalar degrees of
freedom (coupled to the Pontryagin density~\cite{DCS,marco_DCS}) or vector modes (e.g. Einstein-\AE ther theory~\cite{ae_theory}). Note
however that while Einstein-\AE ther theory is classically and quantum mechanically stable, theories 
with pseudoscalars generically present ghosts, unless they are treated as effective field theories~\cite{marco_DCS}.}

To first approximation, we may therefore be tempted to model the equation for linear gravitational perturbations in the odd sector by the Regge-Wheeler equation of GR, but over a generic RZ background metric. This
generalized Regge-Wheeler equation can be  obtained by first writing the spacetime metric as
$g_{\mu\nu}=g^{\rm RZ}_{\mu\nu}+\delta h_{\mu\nu}+{\cal O}(\delta)^2$, with $\delta$  a perturbative
book-keeping parameter and $h_{\mu\nu}$ the metric perturbation. Discarding the ${\cal O}(\delta)^0$ terms of the Einstein equations,
the linear ${\cal O}(\delta)$ terms $\delta R_{\mu\nu}=0$ yield~\cite{hughes}
\begin{equation}\label{eqH}
-\frac12\Box h_{\mu\nu}-\frac12 \nabla_\nu\nabla_\mu h+\nabla_\alpha\nabla_{(\mu} h^\alpha_{\nu)}=0\,,
\end{equation}
with $h=h^\mu_\mu$, $\Box=g_{\rm RZ}^{\mu\nu} \nabla_\mu \nabla_\nu$ and $\nabla$ the covariant derivative defined with the 
background connection.
Assuming then that the metric perturbation has odd parity and adopting the Regge-Wheeler  gauge, i.e. 
\begin{widetext}
\begin{align}
h_{\mu\nu}=&\sum_{lm}h_{\mu\nu,lm} Y^{\ell m}e^{-i\omega t}\,,\\
 h_{\mu\nu,lm}=&
 \begin{pmatrix}
 0&0&-h_0(r)\frac{1}{\sin\theta}\frac{\partial}{\partial\phi}&h_0(r)\sin\theta\frac{\partial}{\partial\theta}\\
 0&0&-h_1(r)\frac{1}{\sin\theta}\frac{\partial}{\partial\phi}&h_1(r)\sin\theta\frac{\partial}{\partial\theta}\\
 -h_0(r)\frac{1}{\sin\theta}\frac{\partial}{\partial\phi}&-h_1(r)\frac{1}{\sin\theta}\frac{\partial}{\partial\phi}&0&0\\
h_0(r)\sin\theta\frac{\partial}{\partial\theta}&h_1(r)\sin\theta\frac{\partial}{\partial\theta}&0&0
 \end{pmatrix}\,,
\end{align}
\end{widetext}
with $Y^{\ell m}$ the spherical harmonics, 
 the same algebraic manipulations that in GR lead to the Regge-Wheeler equation yield 
\begin{align}\label{wave-equation}
\frac{\text{d}^2}{\text{d}{r^{*}}^2}Z + \left[\omega^2 - V_l(r) \right] Z  = 0\,,
\end{align}
 where $Z=N^2 h_1/(r B)$, $\omega$ is the (complex) gravitational wave frequency, the tortoise coordinate $r^{*}$ is related to the areal radius by
\begin{align}
\frac{\text{d} r^{*}}{\text{d}r} = \frac{B(r)}{N^2(r)}\,,
\end{align}
and the potential reads
\begin{equation}\label{eqV}
V_l(r) = \frac{l(l+1)}{r^2}N^2(r) - \frac{3}{r} \frac{\text{d}}{\text{d}r^{*}} \frac{N^2(r)}{B(r)}.
\end{equation}
As can be explicitly verified, the potential reduces to the Regge-Wheeler potential
of GR in the limit in which the RZ metric reduces to Schwarzschild.

Note that in the geometric-optics limit $l\to\infty$,
Eq.~\eqref{eqH} for gravitational perturbations
must reduce to the (null) geodesics equation,
if gravitational waves are to move at the speed
of light (as verified experimentally to within relative errors
of $\sim 10^{-15}$~\cite{GW170817} and as expected if the weak equivalence principle
is to hold). Indeed, this can be seen by
noting that Eq.~\eqref{eqH} becomes $\Box h^{\mu\nu}+2 R^{\mu\alpha\nu\beta}h_{\alpha\beta}=0$ in the Lorenz gauge $h=\nabla_\nu h^{\nu\mu}=0$ (which can
be chosen on any curved vaccum background without loss of generality~\cite{hughes}). One can then insert the ansatz ${h}_{\mu\nu}\approx A_{\mu\nu} \exp(i S)$ into this equation,  keeping only the dominant terms in the limit of large frequencies and wavenumbers ($\partial_\mu S\to\infty$).
This yields the Hamilton-Jacobi equation for massless particles, $g_{\rm RZ}^{\mu\nu} \partial_\mu S \partial_\nu S=0$,
which can be converted explicitly into the null geodesics equation by taking its derivative (see e.g. Sec. 7.8 of Ref.~\cite{defelice} for details).

The fact that gravitational perturbations follow null wavefronts  in the geometric optics limit $l\to\infty$
has important implications for the potential \eqref{eqV}, which 
should necessarily reduce to that of null geodesics  in that limit. Indeed, one
one can easily verify that $V_l(r)\approx \frac{l(l+1)}{r^2}N^2(r)$
for $l\to\infty$, while null geodesics of the RZ metric satisfy
\begin{align}
&\frac{N^4}{E^2} \left(\frac{\text{d}{r^{*}}}{\text{d}{\lambda}}\right)^2+V_r=0\,,\\
&V_r=-1+\frac{b^2 N^2}{r^2}\,,
\end{align}
where $\lambda$  is an affine parameter, $b=L/E$ (with $E$ and $L$ respectively the conserved energy and angular
momentum of the orbit) is the impact parameter, and where we have assumed a reference frame where the orbit is 
equatorial. (This latter assumption  is non-restrictive since we are in spherical symmetry.) As can be seen, the effective 
potential for the radial motion of null geodesics, $V_r$, matches  $V_l\approx {l(l+1)}N^2(r)/r^2$
in the limit $l\sim b\to  \infty$. In particular, 
this implies  that in the geometric optics limit $l\sim b\to  \infty$, the
peak of the effective potential for gravitational perturbations asymptotes to that of the (unstable) circular photon orbit. 
This correspondence in turn implies that at lowest order in the WKB expansion (i.e. in the geometric optics limit),
the real parts of the QNM frequencies are multiples of the orbital frequency of  the circular photon orbit, while
their imaginary parts are related to the Lyapunov exponents of null geodesics near the circular photon orbit (and thus to the curvature of the effective potential
$V_r$ near its peak)~\cite{wkb1,wkb2}.
This can be intuitively interpreted by thinking of QNMs as generated at the circular photon orbit, and slowly leaking outwards 
(since the circular photon orbit is  unstable to radial perturbations).

Two consequences can be drawn from this correspondence between geodesics and gravitational perturbations. To begin with, 
we can conclude that the first term in the effective potential \eqref{eqV} is more robust than the second. Indeed, the first term will
be present in any gravitational theory in which gravitational waves satisfy the equivalence principle and
travel at the speed of  light, as required to high precision by experiments. The second term in Eq. \eqref{eqV} 
is instead less robust, and may depend on the details  of the gravitational theory under scrutiny. 
To check the robustness of our results, we therefore consider also an alternative
phenomenological potential
\begin{equation}\label{eqV2}
V_l(r) = \frac{l(l+1)}{r^2}N^2(r) - \frac{K}{r} \frac{\text{d}}{\text{d}r^{*}} \frac{N^2(r)}{B(r)}\,.
\end{equation}
Note that $K=3$ corresponds to Eq. \eqref{eqV}, while $K=-1$ would correspond to
a scalar field satisfying the wave equation $\Box \phi=0$ on the RZ metric.
However, for generic scalar tensor theories respecting parity, $K$ will be a function of radius, determined by the background metric.\footnote{An example of a theory with $K$ function of $r$ is khronometric theory~\cite{prepFranchini}. In Ref.~\cite{Cardoso:2019mqo} a theory agnostic approach is presented in which the perturbation equations are parametrized.} For simplicity, in what follows we present results for $K=3$ (to be interpreted as a toy model for a situation where the theory of gravity is fixed and thus the equation for the perturbation is known), and for unknown (but constant) $K$. This latter case is a toy model for a situation in which the gravitational theory is unknown. Note that our method allows in principle for a generic function $K(r)$, which we can parametrize by its value and derivatives at the peak of the potential.

One possible future approach to connect parametrized black hole space-times with gravitational field equations has been proposed recently in Ref.~\cite{Suvorov:2020bvk} by building a gravitational theory around the space-time itself. However, connecting this with the Bayesian analysis conducted in this work seems non-trivial, because the underlying theory depends on the background space-time, which itself is varied throughout the analysis.

Furthermore, again in the light of the null geodesics/gravitational waves correspondence, let us note that it would make sense to combine the bounds on the RZ metric from QNM measurements with those coming from observations of the shadow of M87$^*$ by the Event Horizon Telescope (EHT)~\cite{eht,eht2}. We will address this in a forthcoming publication in Ref.~\cite{shadow_paper}.

\subsection{Quasi-Normal Modes}\label{QNM}
Starting from our most general form of the effective potential in Eq. \eqref{eqV2}, it is now our interest to compute the corresponding spectrum of QNM frequencies $\omega_n$. 
To do so we assume the standard black hole boundary conditions that describe purely outgoing waves at spatial infinity and purely ingoing waves at the horizon. 
The QNMs can then be computed by choosing among the many different techniques that have been reported in the literature over several decades.
Detailed information can be found in Refs.~\cite{Kokkotas:1999bd,Nollert_1999,Berti_2009}, which are classical reviews of the field. 
\par 
While the list of methods is long, not all are equally well suited for our application. 
The rather general form of the RZ metric, which has in principle arbitrary many parameters, as well as the computational cost of Bayesian parameter estimation techniques, require an easily adaptable and fast method. 
One such suitable technique is based on the Wentzel-Kramers-Brillouin (WKB) method, which can be used to find approximate solutions to certain types of differential equations \cite{1978amms.book.....B}.
\par 
In the specific context of black hole QNMs  \cite{1985ApJ...291L..33S,1987PhRvD..35.3621I,PhysRevD.35.3632,PhysRevD.37.3378,PhysRevD.41.374,Kokkotas_1991,PhysRevD.68.024018}, the method is well known for providing an approximate solution for the QNM spectrum $\omega_n$.
The method relies only on the knowledge of the Taylor expansion of the effective potential around its maximum and is known to different orders in WKB theory, e.g., the sixth order approximation has been derived in Ref.~\cite{PhysRevD.68.024018}
\begin{align}\label{wkb-expansion}
\frac{i Q_0}{\sqrt{2 Q^{\prime \prime}_0}} - \Lambda_2 - \Lambda_3 - \Lambda_4 - \Lambda_5 - \Lambda_6 = n + \frac{1}{2},
\end{align}
where $Q(r^{*}) \equiv \omega_n^2 - V_l(r^{*})$ is evaluated at the maximum and primes are derivatives with respect to the tortoise coordinate. 
The full expressions of the terms $\Lambda_i$ are rather lengthy, but can be found in the original publication. 
They include higher order derivatives of the potential evaluated at the maximum, as well as the overtone number $n$ under consideration. 
There is no explicit dependency on $l$, because it appears explicitly in $V_l(r)$ itself.
Note that the order of the included derivatives increases by two for every WKB order $i$.
\par 
In this work we follow an approach that allows for a general number of RZ parameters and therefore compute the derivatives numerically with finite differences.
Since the derivatives are taken with respect to the tortoise coordinate, one either has to compute the inverse transformation numerically or compute the derivatives in terms of $r$ or $x$, but then apply the chain rule iteratively.
Because both options can become problematic in terms of precision and computational time for higher order derivatives, especially when the RZ metric has many free parameters, we stop after $\Lambda_2$.
We have verified that the results are very similar to those obtained when $\Lambda_3$ is included as well. 
Because the QNMs used for the parameter estimation are computed with the same method as those for the QNMs we consider as given data, 
we circumvent the problem that WKB is an approximate method. 
\par 
When compared with full numerical results, those of the WKB method are expected to be valid for QNMs with $n < l$, but are less precise and eventually fail for $n \gg l$ (see Ref.~\cite{PhysRevD.68.024018} for a tabulated comparison).
The subsets of QNMs that we consider in this work fall within the valid range $n < l$ of the method. 
Note that another advantage of the WKB method is that one can choose among different orders allowing one to adjust precision and computational cost, which is especially important for a Bayesian analysis.
\subsection{Sets of Models and QNMs}\label{def_spectra}
The most general form of the RZ metric has infinitely many free parameters, which obviously cannot be handled in a numerical approach. 
Therefore, we study different realizations of the RZ metric, in which only a fixed number of free parameters is considered. 
The parameters are not all equally important, as a result of the hierarchical structure of the continued fraction representation. 
Besides the parameters  of the RZ metric, we recall that we have introduced
a parameter $K$ in the potential given by Eq.~\ref{eqV2}.
\par 
In the following, we will consider constraints on several \textit{models}, which
differ by the parameters that we allow to vary. In more detail, we consider the following models:
\begin{align}
\text{model}_{1}  &\equiv \{M, \varepsilon\},					\\
\text{model}_{2}  &\equiv \{M, \varepsilon, a_0, b_0\},			\\
\text{model}_{3}  &\equiv \{M, \varepsilon, a_1, b_1\},			\\
\text{model}_{K1} &\equiv \{M, \varepsilon, K\},				\\
\text{model}_{K2} &\equiv \{M, \varepsilon, a_0, b_0, K\}.		
\end{align}
\par
While the QNM spectrum for each model will in general contain infinitely many modes, any real gravitational wave  experiment can only observe a finite subset of them, see Refs.~\cite{London_2014,Brito_2018,Carullo:2018sfu,PhysRevLett.116.221101,Isi:2019aib,Giesler:2019uxc,Cook:2020otn,Forteza:2020hbw} for recent works on this aspect.
The amplitudes with which QNMs are excited
depend on initial conditions of the black hole perturbations, or on the parameters of the progenitor binary
for QNMs produced after a black hole merger.
The modes that we use in this work correspond to the \textit{typical} QNMs that are excited in the ringdown of binary black hole mergers of comparable mass, see Refs.~\cite{London_2014,Brito_2018,Carullo:2018sfu,PhysRevLett.116.221101,Isi:2019aib,Giesler:2019uxc,Cook:2020otn,Forteza:2020hbw}. 
We consider in particular the Schwarzschild fundamental mode $n=0$ and the first overtone $n=1$, for $l=2$ and $l=3$.
Since whether all four of these modes or only a subset of them can be observed depends on the source signal-to-noise ratio
and  on the gravitational wave detector, we consider two cases (``spectra''), one in which
all four modes are observed, and one in which only the $l=2$ modes are detected. In more detail, we define 
\begin{align}
\text{spectrum}_{1} &\equiv \{l=[2], 		 n=[0,1]		\},	\\
\text{spectrum}_{2} &\equiv \{l=[2, 3], 	 n=[0,1]		\}.
\end{align}
\par
The errors with which we assume that these modes can be measured will be discussed in Sec.~\ref{noise}.
\subsection{Remarks on the RZ Parameter Space}\label{param_space}
While the accuracy of the RZ metric parametrization to describe exact black hole solutions has been studied in several works (e.g. Refs.~\cite{PhysRevD.90.084009,Konoplya:2019fpy,Konoplya:2019goy,konoplya2020general}), using a multi parameter approach for the inverse QNM problem has not been done yet. 
Some single parameter tests using non-QNM data can be found in Ref.~\cite{eht2} using the EHT shadow, or in Ref.~\cite{Cardenas_Avendano_2020} related to using X-ray data and early inspiral gravitational wave information.
In the following we elaborate on two different aspects that one should be aware of when using parametrized metrics for an inverse problem.
\par
The first and more fundamental one is what RZ parameter combinations actually describe black holes.
This is non-trivial to assess if multiple parameters are allowed to vary simultaneously, which could in principle lead to unphysical artifacts. 
As a simple example consider the special case $M=1$ and only $\varepsilon$ as a free parameter.
The requirement that the RZ metric must represent a black hole bounds $-1 < \varepsilon \leq 1/2$~\cite{PC}.
Similarly, when 
$\{\varepsilon,a_0,b_0\}$ or $\{\varepsilon,a_1,b_1\}$ are allowed to vary, with the other parameters set to zero,
this constraint becomes  $-1 < \varepsilon \leq  (1+a_0)/2$ or $-1 < \varepsilon \leq  (1+a_1)/2$]~\cite{PC}.
\par
The second aspect is related to our  choice of using the higher order WKB method.
It is a priori not clear what combinations of RZ parameters only lead to small deformations of the Regge-Wheeler potential, and what combinations describe 
instead large and qualitative differences, e.g. regions where the potential is negative.
The latter case would question the validity of the higher order WKB method, and might lead to bound states.
To be sure that the method is justified, one has to quantify the \textit{allowed regions} of the parameter space, which can in principle be used as priors for the Bayesian parameter estimation that we will undertake below.
\par
We attempted to tackle this issue  by  sampling the parameter space by brute force,
checking at each point if the potentials becomes negative somewhere.
However, the number of total computations $N_\text{total}$ needed for a single choice of $l$ scales as
\begin{align}
N_\text{total} \propto N_\text{res-pot} \times \left( {N_\text{res-param}} \right)^{D},
\end{align}
where $N_\text{res-pot}$ and $N_\text{res-param}$ are the number of sampling points for
the potential and for each  of the $D$ parameters. Therefore, it is evident that the problem becomes easily unmanageable from a computational point of view.
In practice, however,  one already knows that the parameters describing the Schwarzschild limit are allowed, and 
can  start by sampling the parameter space around this limit, then progressively moving away from it. 

Since the mass of the final black hole remnant is expected to be within 5--10\% of 
the mass of the progenitor binary\footnote{Note that in
GR, if the  masses of the progenitor black holes are known, the remnant's mass is also known~\cite{morozova}. However, here
we obviously cannot rely on GR, since our goal is to test it. Nevertheless, since 5--10\% is the typical mass loss due to
gravitational wave emission in GR~\cite{morozova}, it seems reasonable to assume that the final mass will be known to within at least that error, also beyond GR.}, in the following we consider $M$ to be within $[0.9, 1.1]$ and vary different RZ parameters.
From our numerical analysis it seems that the $l=2$ potentials are more prone to becoming negative.
\par
\begin{figure}
	\centering
	\includegraphics[width=1.0\linewidth]{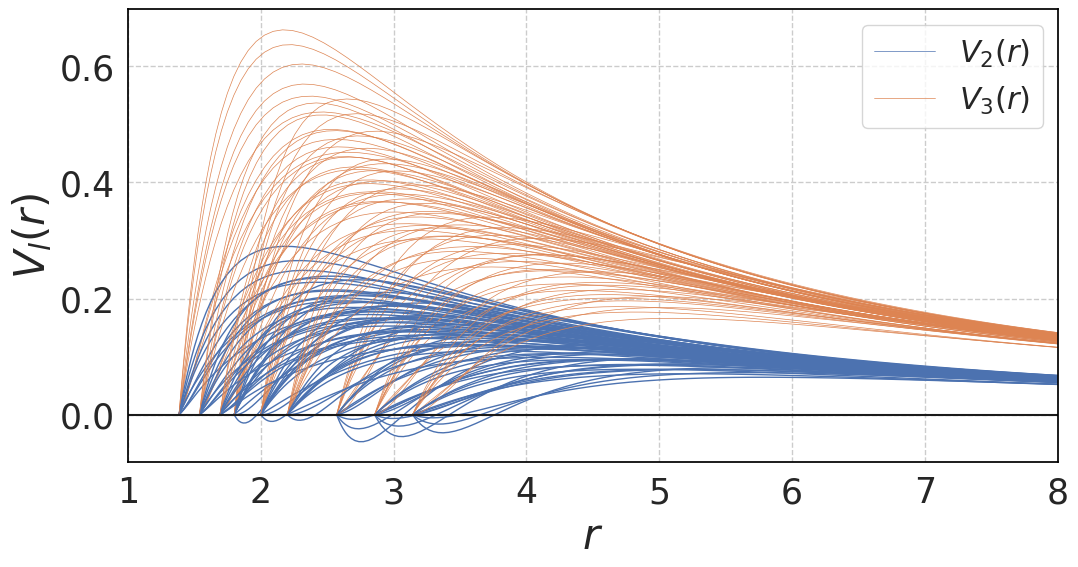}
	\caption{Potentials for $M$ in the range $[0.9, 1.1]$ and for the RZ parameters $\{\varepsilon, a_0, b_0\}$ in the range $[-0.3, 0.3]$. \label{potential_region_Mea0b0_positive-negative}}
\end{figure}

 
For the simplest model$_1$, we verified that $\varepsilon$ within $[-0.5, 0.5]$ does not lead to negative potentials.
For model$_2$ we find that ranges of $[-0.15, 0.15]$ for $\{\varepsilon, a_0, b_0\}$ are fine, but extending them further starts  becoming problematic (i.e.
some combinations become invalid).
In Fig.~\ref{potential_region_Mea0b0_positive-negative} we show a sample of the potentials where the RZ parameters $\{\varepsilon, a_0, b_0\}$ are within $[-0.3, 0.3]$, and include a few of such invalid combinations.
Note that most of the potentials are positive everywhere, but some become negative close to the horizon.
\par
This observation might seem troublesome for parameter estimation, unless priors are chosen such that
negative potentials are excluded  from the start. 
However, in practice, negative potentials are not only rare, but also tend to
produce large deviations from GR in the QNMs computed with our WKB approximation. As  a result,
even if parameters producing negative potentials are drawn during sampling, they will almost always be rejected.
For the cases presented in this work, we explicitly show that the sampled parameter combinations do not include negative potentials or other possible large deviations in Sec.~\ref{applications}.

Also note that even if certain set of parameters may in principle produce large deviations of the potential near the horizon, the early time-evolution of perturbations would still contain modes similar to the QNMs computed with a WKB approximation at the potential's peak, while the true QNM spectrum might be different, but appears at later times. 
This has been first studied for ultra compact stars in Ref.~\cite{1996gr.qc.....3024K} and further pursued in Refs.~\cite{PhysRevD.60.024004,2000PhRvD..62j7504F}. Nowadays this phenomenon related to the leakage of trapped $w$-modes is known as ``echoes'' and applies to exotic compact objects~\cite{Barausse:2014tra,Cardoso:2016rao}, some types of wormholes \cite{Bueno:2017hyj,paper5} and various types of modifications on the horizon scale and phenomenological models of quantum black holes \cite{2017PhRvD..96h2004A,Maggio:2017ivp,Nakano:2017fvh,Barcelo2017,Wang:2019rcf,Oshita:2019sat,Cardoso:2019apo,paper9}.
\subsection{Treatment of Noise}\label{noise}
While one can obtain the full QNM spectrum exactly (up to modeling errors due to the WKB approximation
and to numerical errors), real observations will always present a certain degree of uncertainty. 
The details of the latter will depend on the specific source parameters, as well as on the properties of the detector and on the data analysis technique 
used to extract the signal and estimate its parameters~\cite{Berti:2016lat}. 
Since this is a major problem in itself, we have adopted here a simplified approach. 

First, we treated observational errors  in the reconstructed spectrum by adding a Gaussian noise to our theoretically computed QNM frequencies and decay times. 
This produces an intrinsic variation in the reconstructed parameters, because every realization of the noise is unique. This is especially problematic because of
the relatively small number of modes that we use as data. 
To account for this bias due to the realization of the errors, one would have to repeat each MCMC analysis for many different realizations of them.
While this analysis is beyond the scope of this work, we have explored several realizations to make sure that our parameter reconstruction works correctly.
However, for the rest of this work we will adopt the  noiseless limit, i.e. we inject the exact QNMs as input data for the parameter estimation. 

As for the  variance of our Gaussian noise, we consider two possibilities.
To mimic the error on the measured QNMs that would be achieved with
Advanced LIGO and Virgo at design sensitivity and events similar to GW150914, we assume that 
QNMs are known within  $1\sigma$ relative uncertainties of about $10\,\%$: See e.g. Fig. 5 of Ref.~\cite{PhysRevLett.116.221101}, and Figs. 2 and 4 of Ref.~\cite{Isi:2019aib} for uncertainties on the measured QNMs with O1 data; Refs. \cite{Brito_2018,Carullo:2018sfu,Giesler:2019uxc,Cook:2020otn,Forteza:2020hbw} for 
reports on the simultaneous extraction of several QNMs from numerical relativity simulations; and 
Ref. \cite{Yang:2017zxs,Maselli:2017kvl} for the possibility of stacking several modes together to enhance tests of the no-hair theorem. Furthermore, we consider $1\sigma$ relative uncertainties of $ 1\,\%$ to mimic next generation detectors like the Einstein Telescope~\cite{ET} or LISA~\cite{lisa}, or especially loud events~\cite{Berti:2016lat}.
\subsection{Markov chain Monte Carlo}\label{pymc3}
Our Bayesian parameter estimation pipeline relies on Markov chain Monte Carlo (MCMC) techniques. 
This class of methods allows for sampling the posterior distribution of the parameters of a model that is used to describe a given set of data.
Since a detailed introduction to Bayesian analysis and MCMC methods is beyond the scope of this work, we only summarize here the key aspects of our framework and refer the interested reader to Ref.~\cite{10.5555/971143} for a comprehensive introduction.
\par 
To perform the MCMC analysis we utilize the
Metropolis Hastings sampler of the Python based probabilistic programming framework \textsc{PyMC3} \cite{pymc3},
which we couple (via a custom theano function) to an external \textsc{C++} code computing the potentials. To enhance the computational performance, we initialize 6 chains that are computed in parallel (12 for model$_3$ and model$_{K2}$). Furthermore, we set 10k tuning steps in the \textsc{PyMC3} subroutine to optimize the sampling, which are discarded from the analysis. Depending on the specific model, we remove at least the first 10k steps in each chain for burn in. In each chain there are at least 100k steps for the simple models and up to 2000k for the most complex one.
Depending on the choice of the model, the total number of steps and the provided QNMs, one analysis will typically take between several minutes to a few hours.
\par 
Our likelihood follows from our simplistic assumption that the measured QNMs are affected
by Gaussian errors. In more detail, we write the likelihood as a product
of Gaussians for the real and imaginary parts of the QNMs that we assume
are measured, centered on the true values of the modes and with 
standard deviation corresponding, as discussed above, to $10\,\%$ or $1\,\%$
of the true values.
\par 
Bayesian analysis also requires one to specify priors for the parameters, which reflect our knowledge on them before looking at the data. 
We assume a Gaussian prior on the black hole mass $M$.
As mentioned earlier, in GR the final mass is lower than the initial
total binary mass by 5-10\%~\cite{morozova} due to the emission of gravitational waves. The final
value of the mass can be computed from the initial mass of the binary via
simple formulae in GR~\cite{morozova}, but similar formulae do not yet exist 
in modified gravitational theories. Nevertheless, one would expect GW energy losses beyond GR 
to be of the same order of magnitude as in GR, which would make the final
mass known (because coinciding with the binary's initial mass) up to a 5-10\% uncertainty. Moreover,
this prior can be improved simply by calculating the energy flux in the gravitational wave signal (a calculation
that can be performed from the data alone, irrespective of the theory). To account for this additional information,
we choose a standard deviation of $2.5\%$ around the real value ($M=1$) for our Gaussian prior on the mass.
We will see in the following that this prior is rather uninformative, at least for models 
with few parameters and for precise QNM measurements, i.e. the posteriors are dominated by the likelihood and not by the prior.
\par 
From our discussion of the RZ parameter space in Sec.~\ref{param_space}, we know that the other prior ranges also have to be chosen with caution.
For all RZ parameters, $a_0$, $b_0$, $a_1$, $b_1$ and $\varepsilon$, we adopt flat  priors centered on the Schwarzschild values  $a_0=b_0=a_1=b_1=\varepsilon=0$ 
and with width of $\pm 1$. This width is motivated by the fact that one expects
values of these dimensionless parameters differing from GR by more than ${\cal O}(1)$
should be disfavored by other observations, and particularly GW observations of
the inspiral of BH binaries~\cite{Cardenas_Avendano_2020} (especially with future detectors, see e.g.~\cite{Barausse:2016eii}). For $K$, for similar reasons we adopt a flat prior centered on the GR value $K=3$, and width of $\pm 5$.
While these priors may contain parameters combinations for which the WKB approximation breaks down, we have verified a posteriori that 
the sampling chains tend to avoid those combinations. The robustness of our results with respect to the choice of priors is  discussed in detail in Sec.~\ref{discussion-priors}.

\section{Applications}\label{applications}
\begin{figure*}
	\centering
	\begin{minipage}{0.98\linewidth}
		\begin{minipage}{0.5\linewidth}
		\centering
		\includegraphics[width=1.0\linewidth]{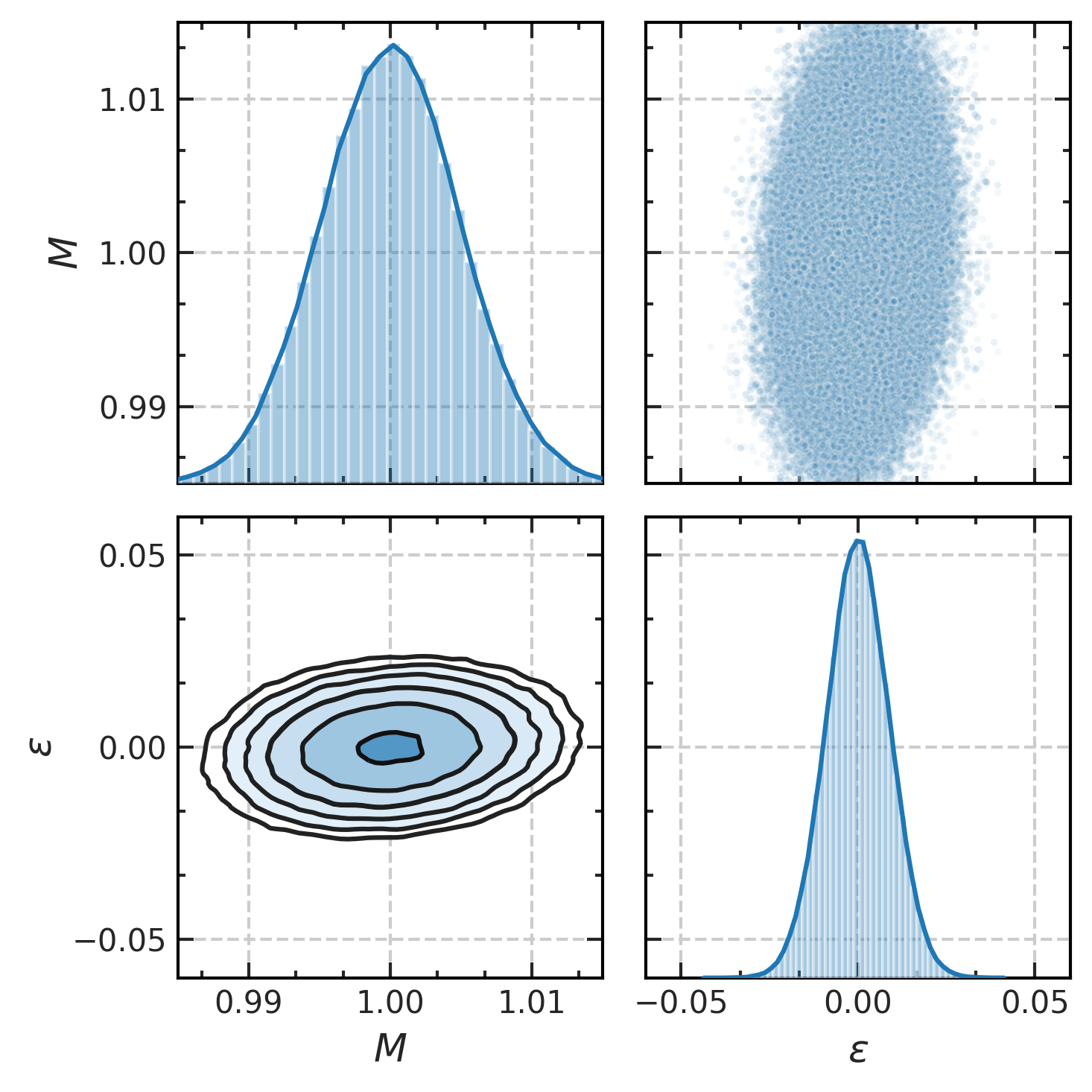}
		\end{minipage}
	\begin{minipage}{0.45\linewidth}
		\begin{minipage}{1.0\linewidth}
				\vfill
			\centering
			\includegraphics[width=1.0\linewidth]{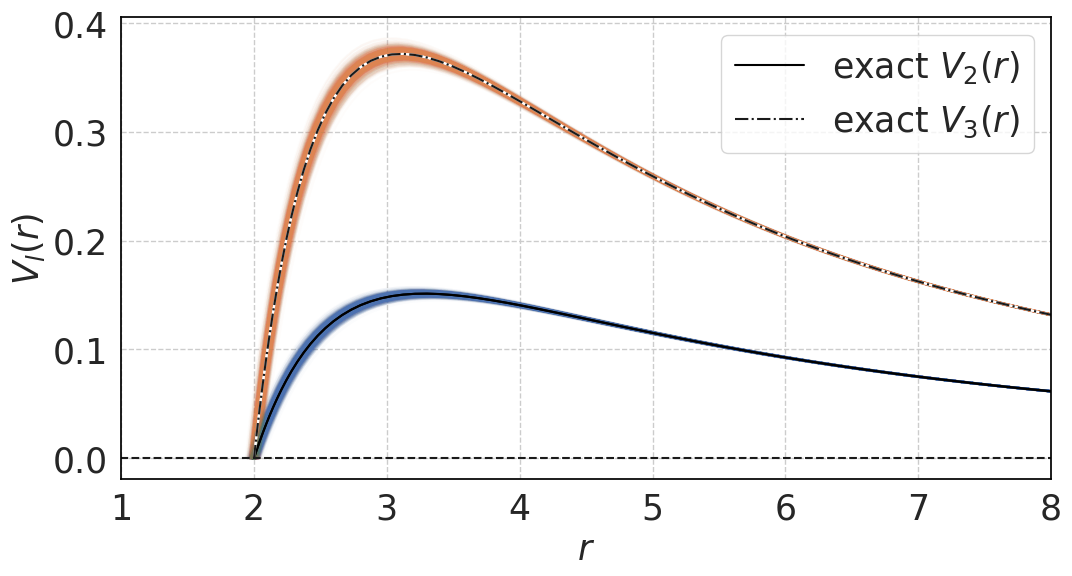}
		\end{minipage}
		\vfill
		\begin{minipage}{1.0\linewidth}
			\centering
			\includegraphics[width=1.0\linewidth]{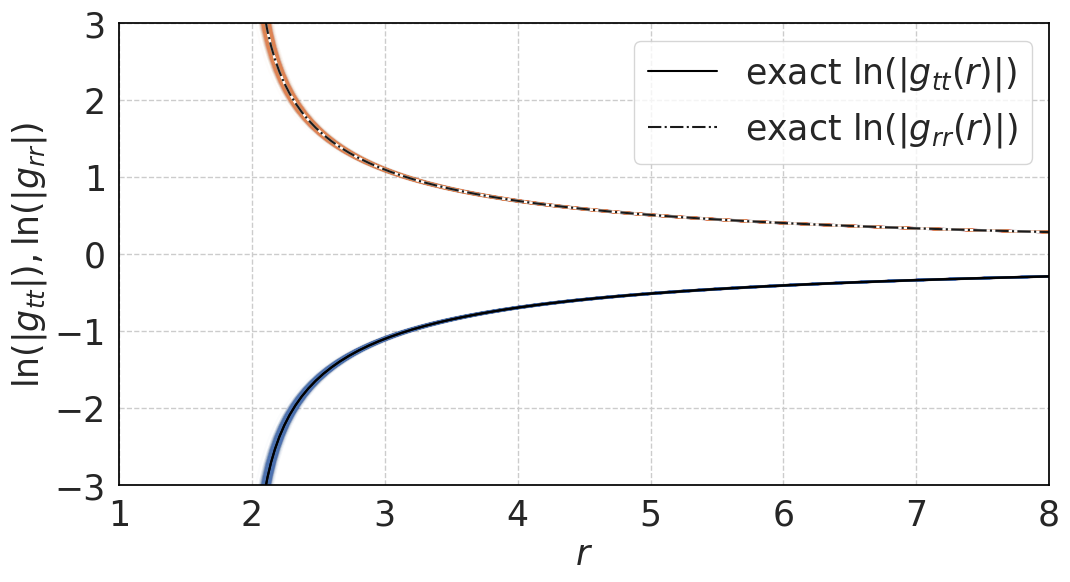}
		\end{minipage}
				\vfill
	\end{minipage}
	\end{minipage}
	\caption{Results for model$_1$ obtained by using spectrum$_1$ with $\pm 1\,\%$ relative errors. 
			\text{Left:} MCMC parameter estimation.
			\text{Right top:} Exact (black lines) and reconstructed (colored lines) potentials $V_2(r)$ and $V_3(r)$.
			\text{Right bottom:} Exact (black lines) and reconstructed (colored lines) metric functions $g_{tt}(r)$ and $g_{rr}(r)$.
			\label{fig_modelMe}
			}
\end{figure*}

\begin{figure*}
	\centering
	\begin{minipage}{0.98\linewidth}
		\begin{minipage}{0.5\linewidth}
		\includegraphics[width=1.0\linewidth]{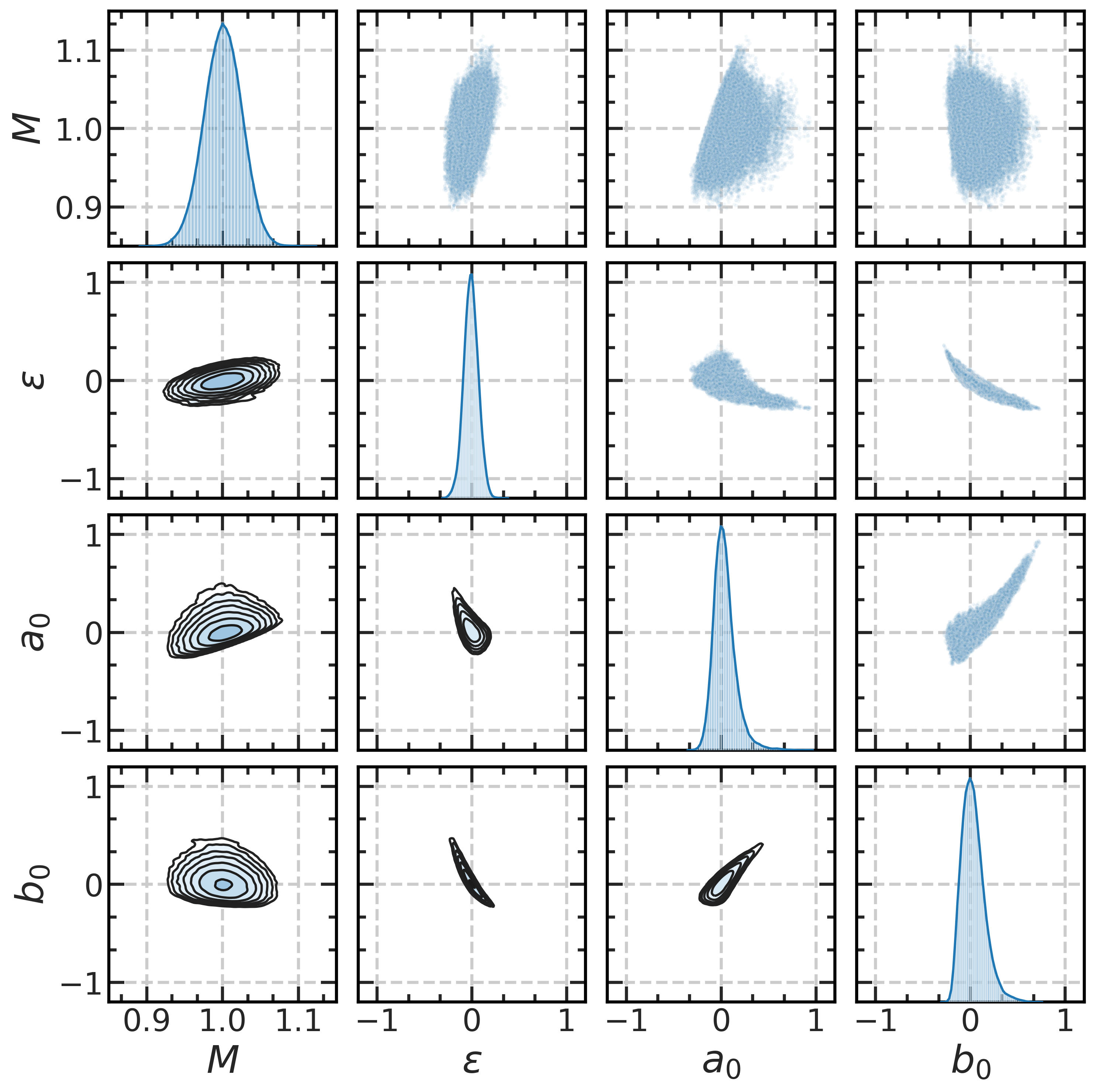}
		\end{minipage}
	\hfill
	\begin{minipage}{0.455\linewidth}
		\begin{minipage}{1.0\linewidth}
			\centering
			\includegraphics[width=1.0\linewidth]{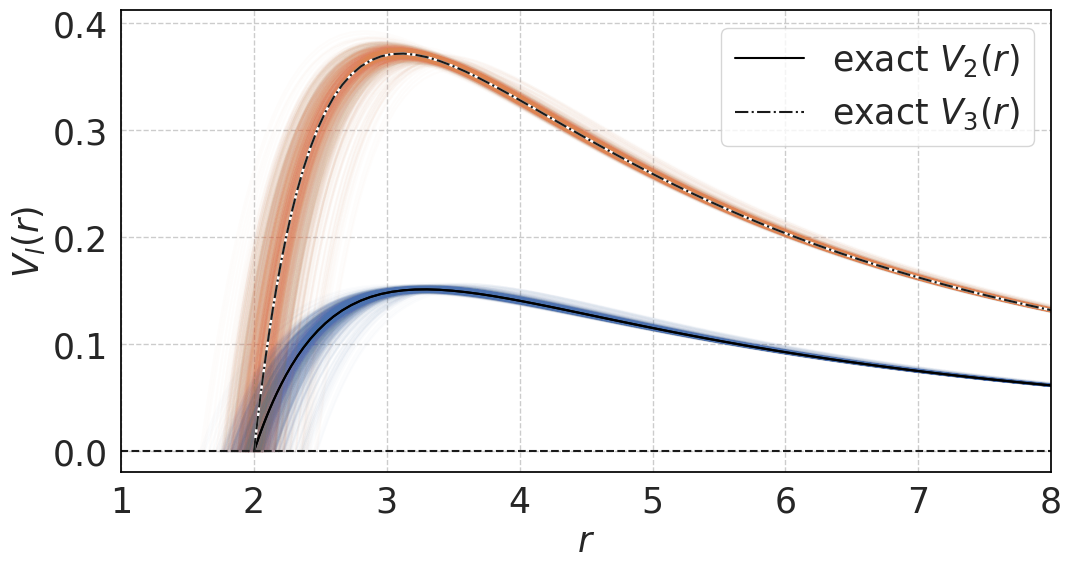}
		\end{minipage}
		\vfill
		\begin{minipage}{1.0\linewidth}
			\centering
			\includegraphics[width=1.0\linewidth]{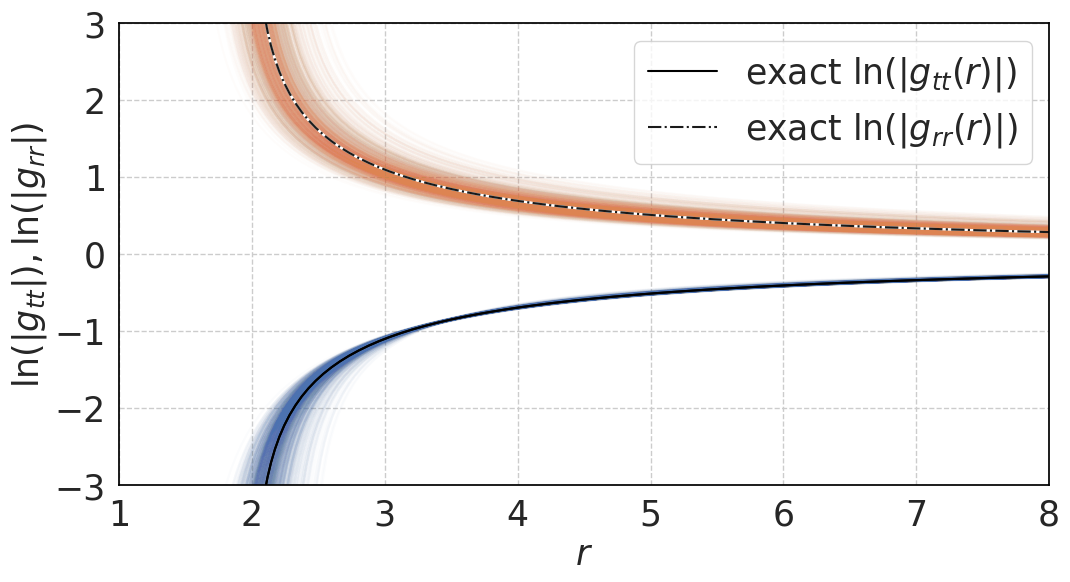}
		\end{minipage}
	\end{minipage}
	\end{minipage}
	\caption{Results for model$_2$ obtained by using spectrum$_2$ with $\pm 1\,\%$ relative errors. 
	\text{Left:} MCMC parameter estimation.
	\text{Right top:} Exact (black lines) and reconstructed (colored lines) potentials $V_2(r)$ and $V_3(r)$.
	\text{Right bottom:} Exact (black lines) and reconstructed (colored lines) metric functions $g_{tt}(r)$ and $g_{rr}(r)$.
		\label{fig_modelMea0b0}}
\end{figure*}

\begin{figure*}
	\centering
	\begin{minipage}{0.98\linewidth}
		\begin{minipage}{0.5\linewidth}
		\includegraphics[width=1.0\linewidth]{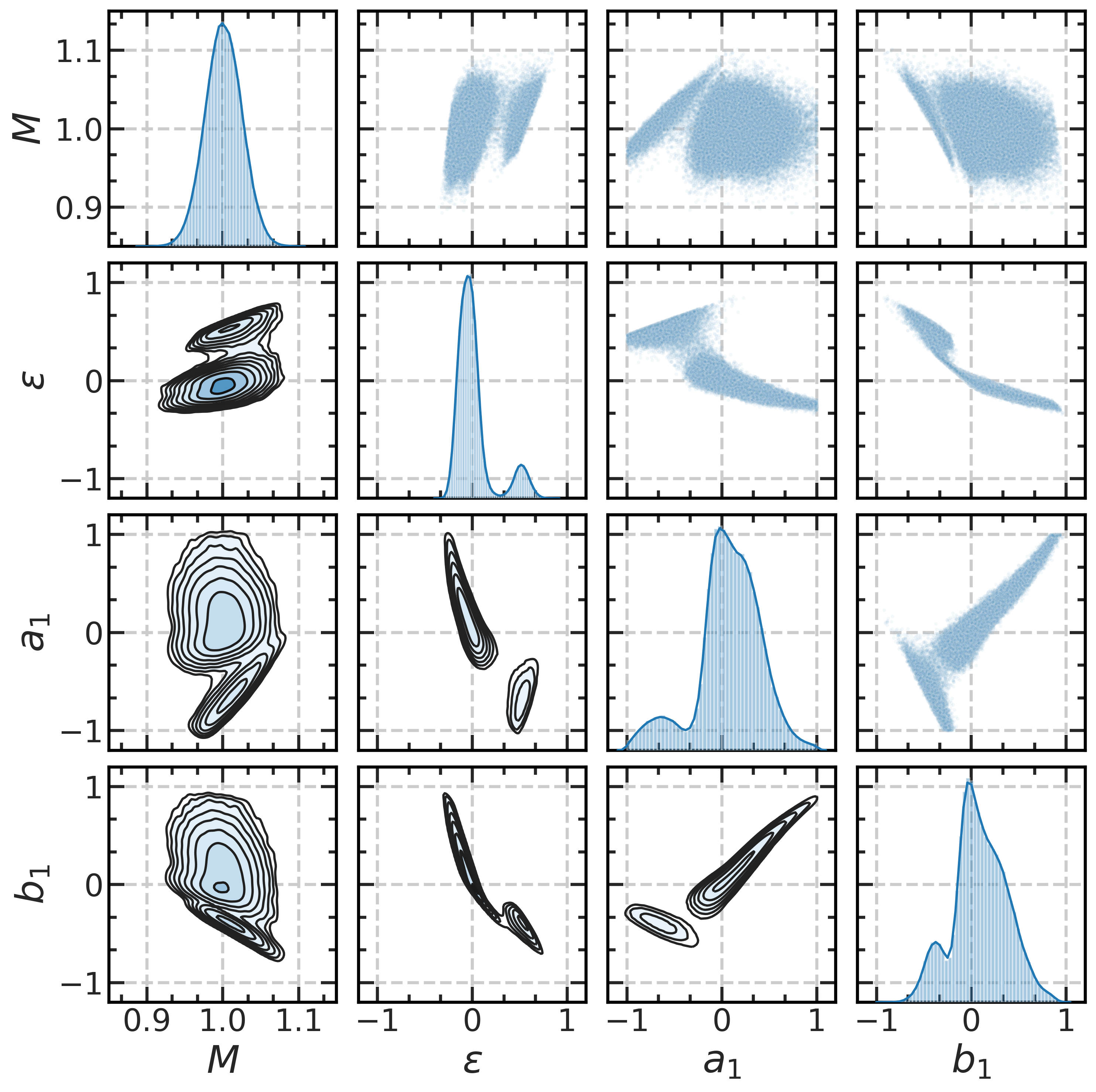}
		\end{minipage}
		\hfill
		\begin{minipage}{0.455\linewidth}
		\begin{minipage}{1.0\linewidth}
			\centering
			\includegraphics[width=1.0\linewidth]{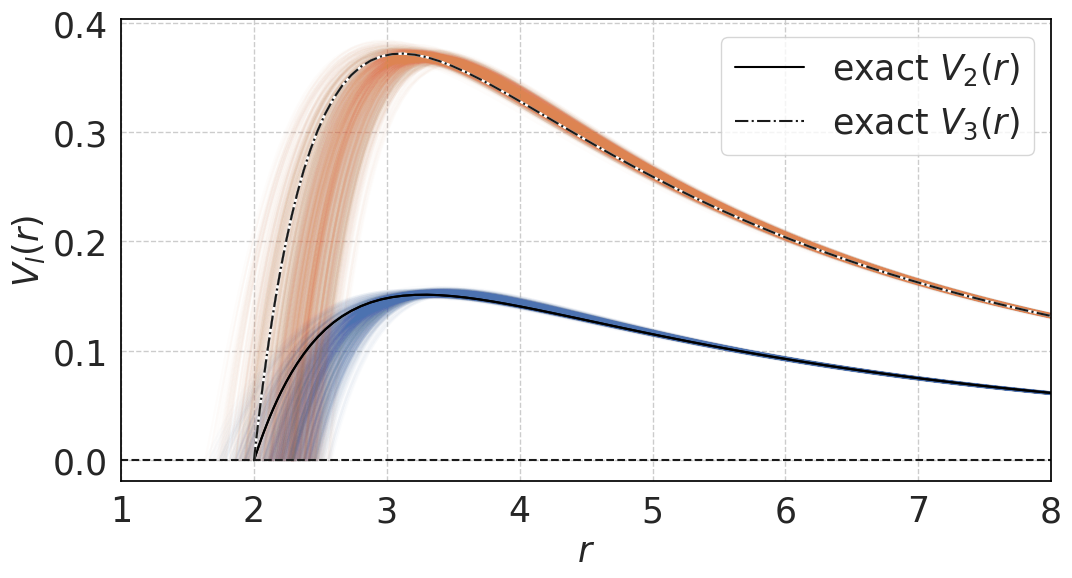}
		\end{minipage}
		\vfill
		\begin{minipage}{1.0\linewidth}
			\centering
			\includegraphics[width=1.0\linewidth]{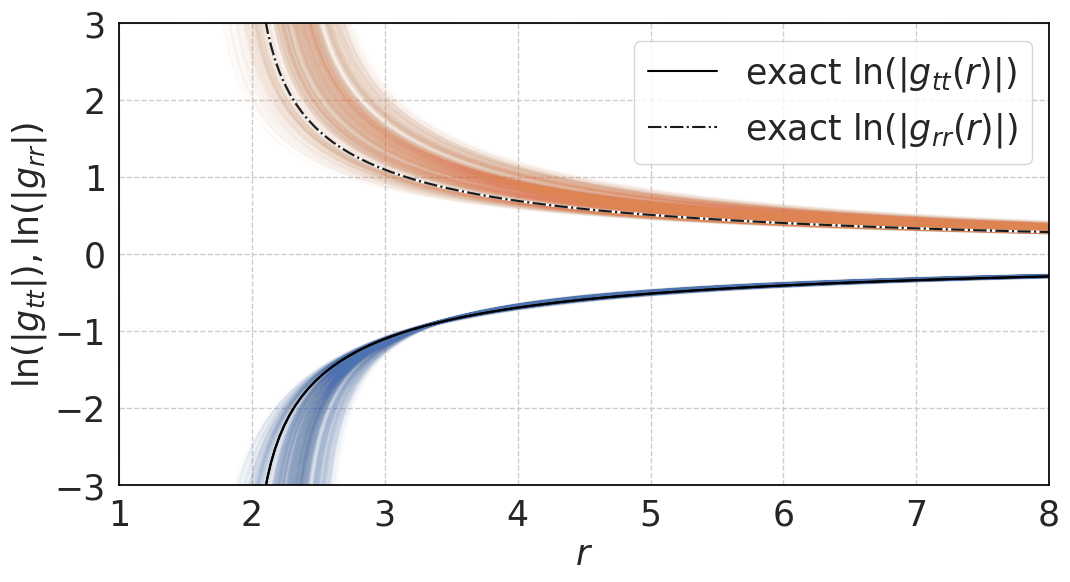}
		\end{minipage}
	\end{minipage}
	\end{minipage}
	\caption{Results for model$_3$ obtained by using spectrum$_2$ with $\pm 1\,\%$ relative errors. 
	\text{Left:} MCMC parameter estimation.
	\text{Right top:} Exact (black lines) and reconstructed (colored lines) potentials $V_2(r)$ and $V_3(r)$.
	\text{Right bottom:} Exact (black lines) and reconstructed (colored lines) metric functions $g_{tt}(r)$ and $g_{rr}(r)$.
		\label{fig_modelMea1b1}}
\end{figure*}

\begin{figure*}
	\centering
	\begin{minipage}{0.98\linewidth}
		\begin{minipage}{0.5\linewidth}
		\includegraphics[width=1.0\linewidth]{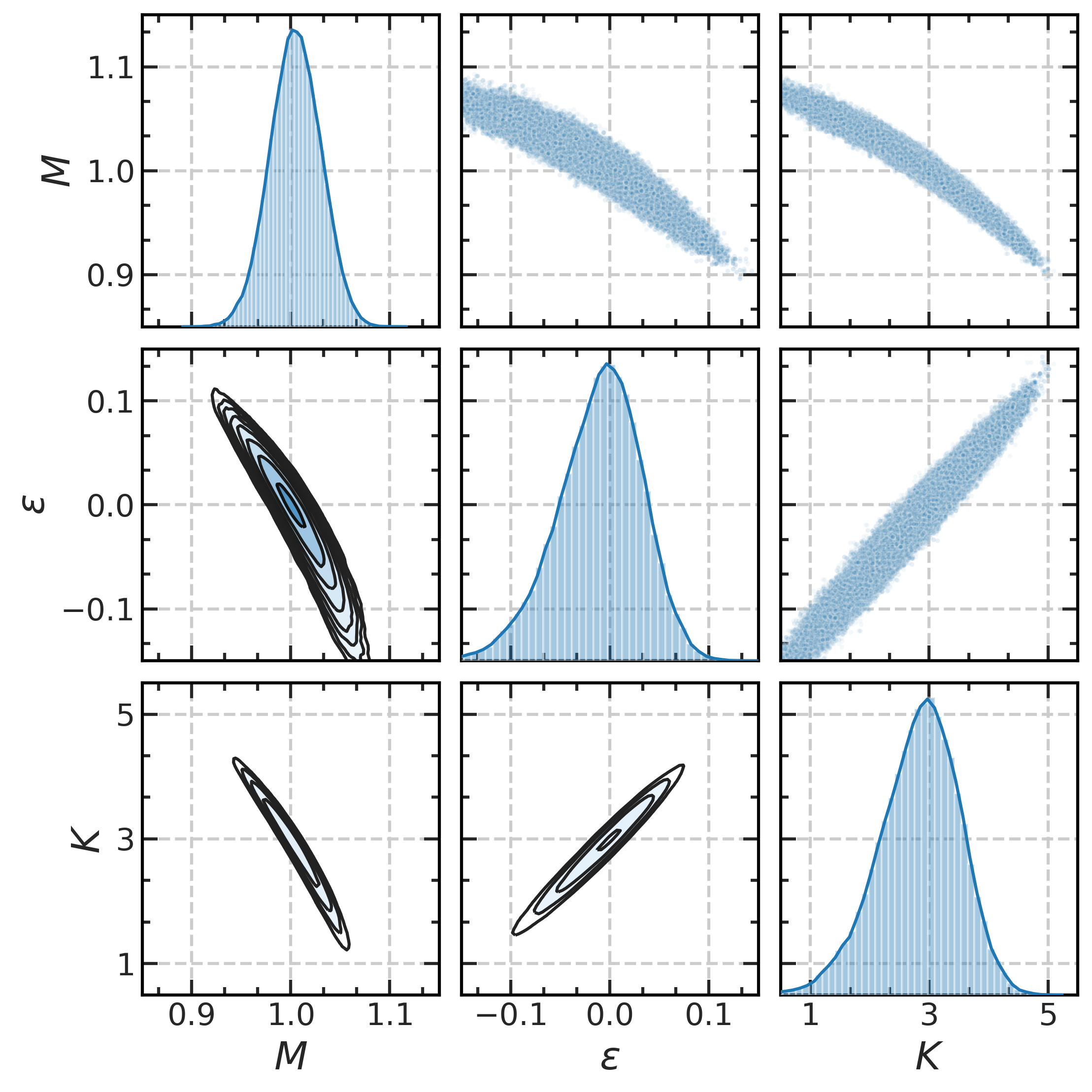}
		\end{minipage}
		\hfill
		\begin{minipage}{0.455\linewidth}
		\begin{minipage}{1.0\linewidth}
			\centering
			\includegraphics[width=1.0\linewidth]{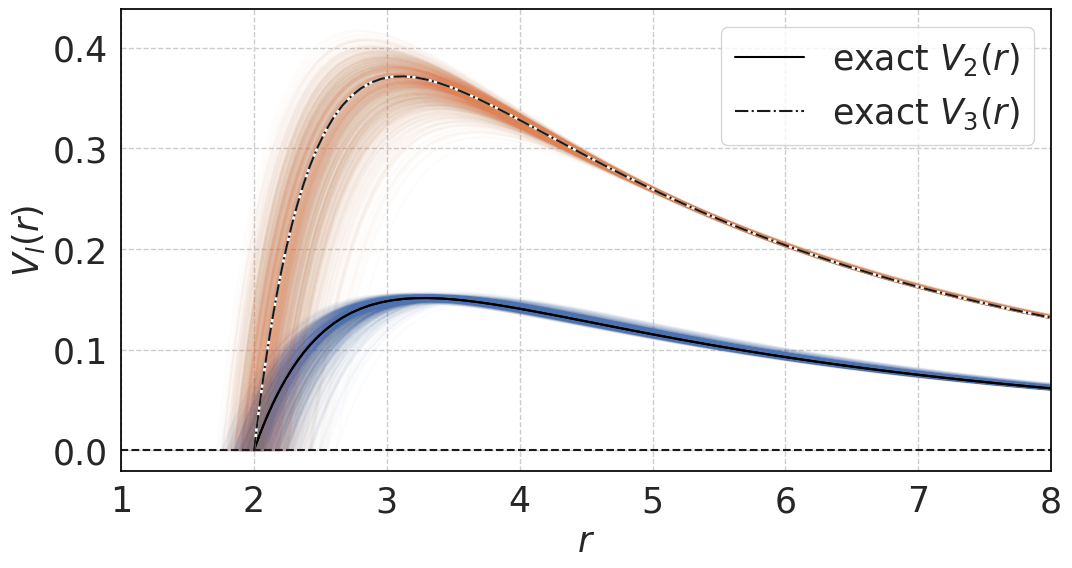}
		\end{minipage}
		\vfill
		\begin{minipage}{1.0\linewidth}
			\centering
			\includegraphics[width=1.0\linewidth]{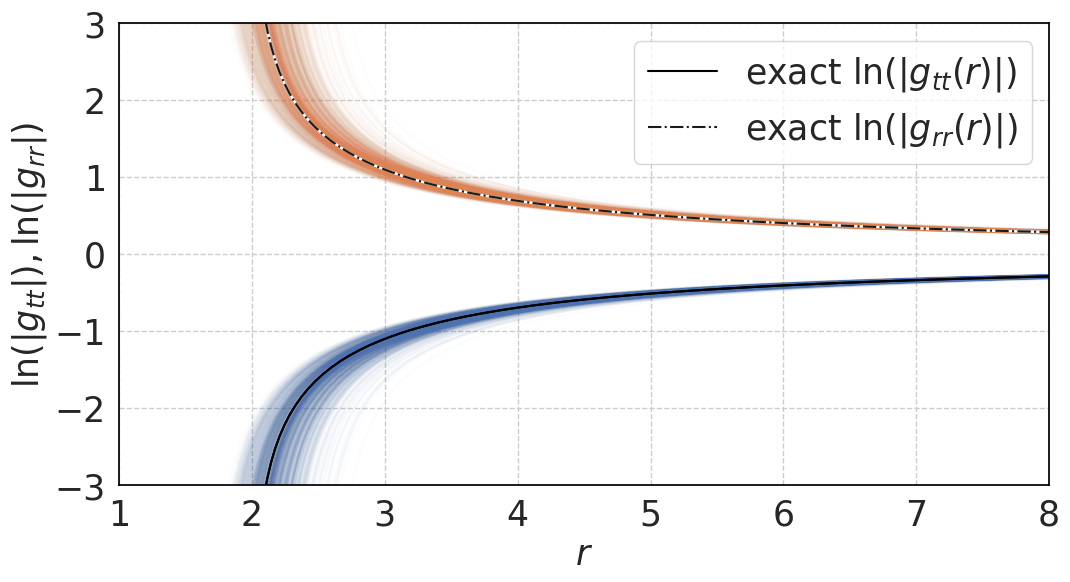}
		\end{minipage}
	\end{minipage}
	\end{minipage}
	\caption{Results for model$_{K1}$ obtained by using spectrum$_1$ with $\pm 1\,\%$ relative errors. 
	\text{Left:} MCMC parameter estimation.
	\text{Right top:} Exact (black lines) and reconstructed (colored lines) potentials $V_2(r)$ and $V_3(r)$.
	\text{Right bottom:} Exact (black lines) and reconstructed (colored lines) metric functions $g_{tt}(r)$ and $g_{rr}(r)$.
		\label{fig_modelMeK}}
\end{figure*}

\begin{figure*}
	\centering
			\includegraphics[width=0.8\linewidth]{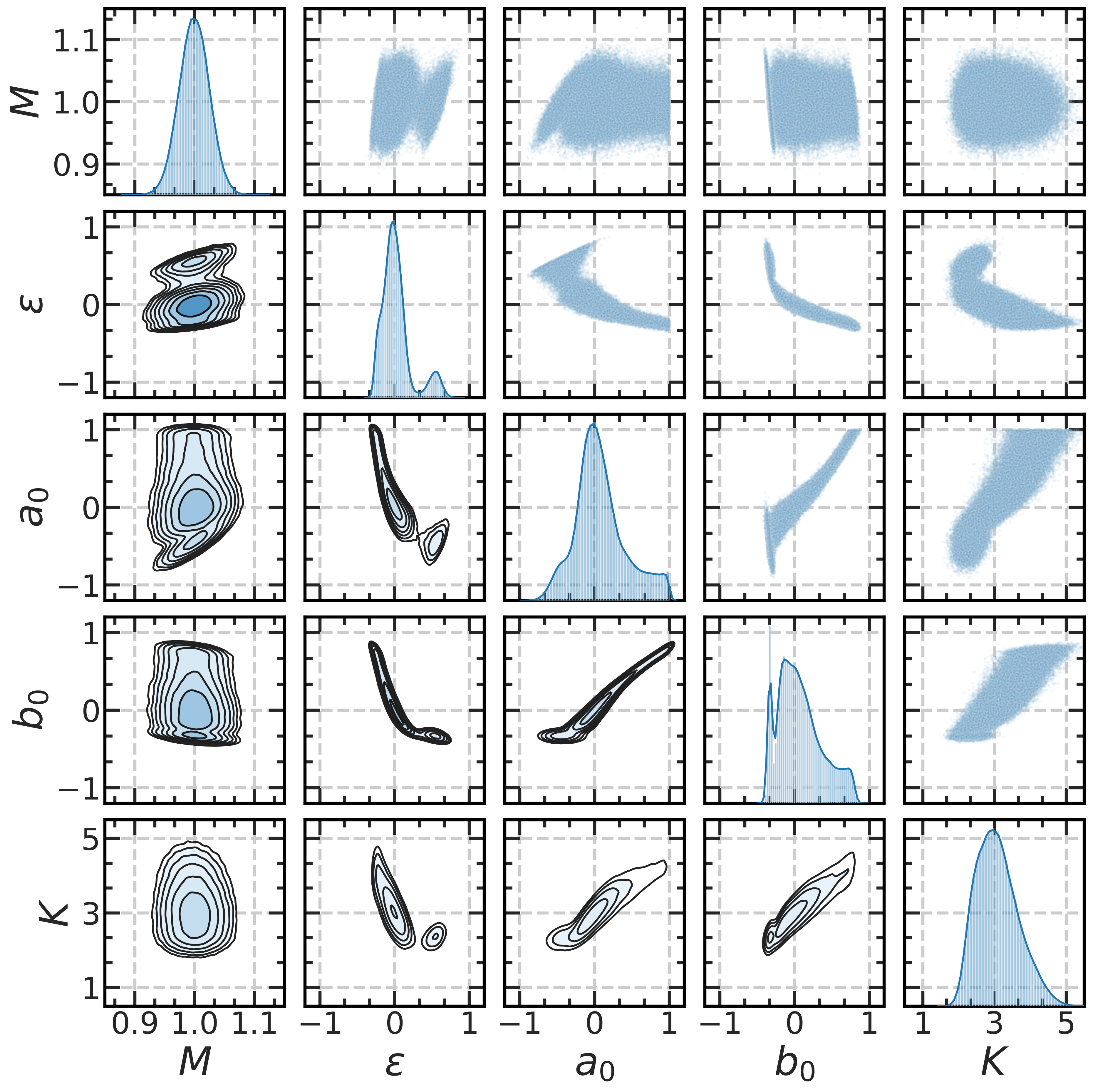}
		\caption{MCMC parameter estimation for model$_{K2}$ obtained by using spectrum$_2$ with $\pm 1\,\%$ relative errors. 
		\label{fig_modelMea0b0K_1}}
		\end{figure*}

\begin{figure}
			\begin{minipage}{1.0\linewidth}
				\centering
				\includegraphics[width=1.0\linewidth]{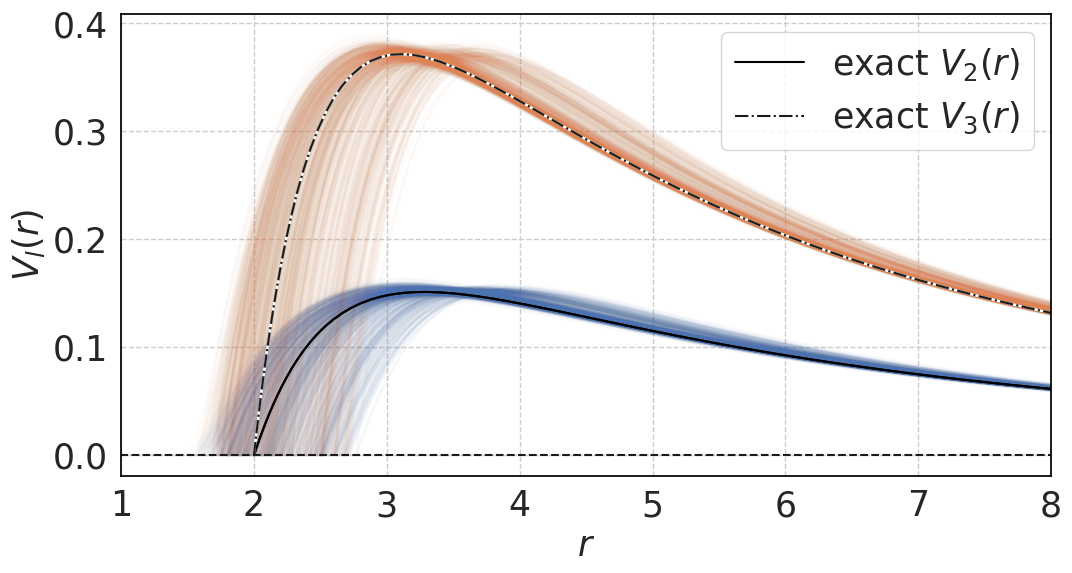}
			\end{minipage}
			\vfill
			\begin{minipage}{1.0\linewidth}
				\centering
				\includegraphics[width=1.0\linewidth]{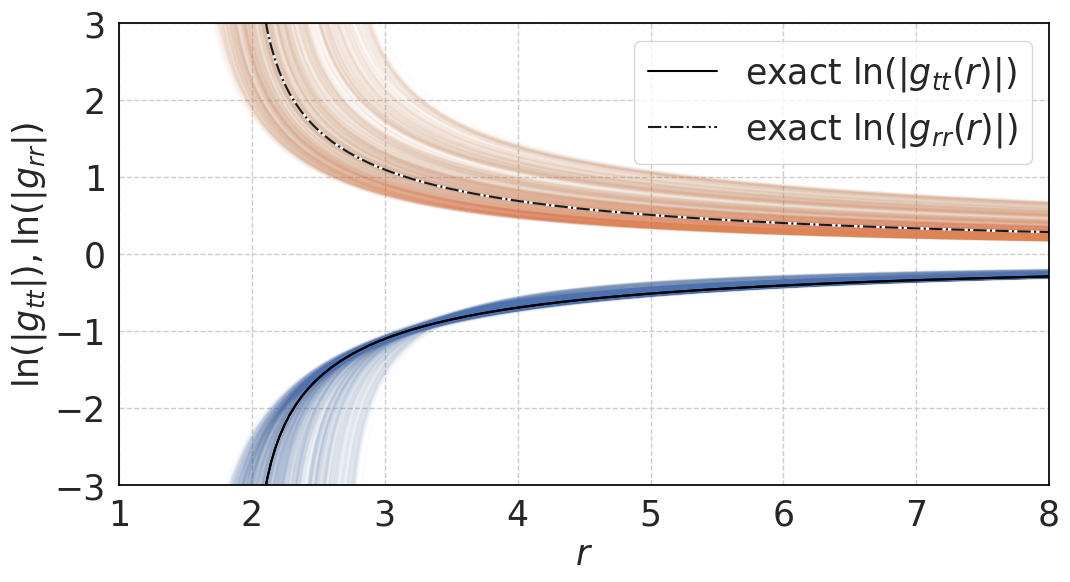}
			\end{minipage}
	\caption{Results for model$_{K2}$ obtained by using spectrum$_2$ with $\pm 1\,\%$ relative errors. \text{Top:} Exact (black lines) and reconstructed (colored lines) potentials $V_2(r)$ and $V_3(r)$. \text{Bottom:} Exact (black lines) and recon- structed (colored lines) metric functions $g_{tt}(r)$ and $g_{rr}(r)$.
		\label{fig_modelMea0b0K_2}}
\end{figure}

In this section, we apply our methods to the different models outlined above, and show representative examples of our results in Figs.~\ref{fig_modelMe}, \ref{fig_modelMea0b0}, \ref{fig_modelMea1b1}, \ref{fig_modelMeK}, \ref{fig_modelMea0b0K_1} and \ref{fig_modelMea0b0K_2}.
Each figure summarizes the MCMC parameter estimation, the reconstructed potentials and the reconstructed metric functions for a specific model with given QNM spectrum and QNM measurement errors, as described in the caption \footnote{Note that we split the left and right panel figure layout for model$_{K2}$ into two separate figures for better readability}.
Although we have studied all combinations of the two different QNM subsets with the two different assumptions for the 
measurement errors for each of the five different models (i.e. a total of 20 combinations), we
do not show all of them here for reasons of space, and because the results for most models scale roughly linearly 
with the assumed errors of the QNM measurements (once the spectrum and the model are fixed). 
The structure of the figures is the same for most models and described in the following.
\par 
Each of the MCMC parameter estimation results is shown on the left panel.
There, the diagonal sub-panels show the posterior distributions for the free parameters of the model under investigation. 
The sub-panels above the diagonal are scatter plots in which each point stands for one step of the chain. 
Because our chains contain around one million steps, we also show the corresponding contour plots in the sub-panels below the diagonal.
Adjacent contours correspond to values differing by 0.3 dex in logarithmic scale (i.e. by a factor 2).
\par
In all panels related to the potentials and metric functions we provide the exact injected functions in solid and dashed black lines, while colored lines show the reconstruction.
In order to visualize and quantify the uncertainties coming from the parameter estimation, we draw 1000 random samples from the MCMC chains and evaluate the potentials and metric functions for those parameters. 
These are then added as semi-transparent colored lines, which make regions of high confidence appear more saturated. 
Note that the potentials are always shown for $l=2$ and $l=3$, even when the QNM measured spectrum does not contain $l=3$. 
In this case, we compute the $l=3$ potential from the reconstructed parameters from the $l=2$ spectrum, in order to see how it compares with the reconstructed spectrum when $l=3$ QNMs are included.


\section{Discussion}\label{discussion}
In this section we discuss our findings, starting with the MCMC parameter estimation and related details in Sec.~\ref{discussion-parameter-estimation}, while the reconstructed potentials and metric functions are addressed in Sec.~\ref{discussion-potential-metric}.
More details on the results that we obtain when solar system PPN bounds are imposed as priors are presented in Sec.~\ref{discussion-ppn}.
We briefly comment on other recent works that use QNMs for the inverse problem in Sec.~\ref{discussion-alt-inv}. 
Finally we discuss possible extensions of this work in Sec.~\ref{discussion-extensions}.
\subsection{Parameter Estimation}\label{discussion-parameter-estimation}
From the MCMC results presented in the left panels of Figs.~\ref{fig_modelMe}, \ref{fig_modelMea0b0}, \ref{fig_modelMea1b1} and \ref{fig_modelMeK}, as well as in Fig.~\ref{fig_modelMea0b0K_1}, one can see that for all models one can put constraints on all of the parameters, though there are great differences between them.
\par
In general, the posteriors obtained for the low-dimensional models, which include $\varepsilon$ as  the only free parameter, are usually more constrained than those of the higher-dimensional models.
For example, using spectrum$_1$ for model$_1$ and model$_{K1}$ (with results shown in Fig.~\ref{fig_modelMe} and Fig.~\ref{fig_modelMeK}), one finds that $M$ and $\varepsilon$ can in both cases be well constrained, but the additional parameter $K$ clearly impacts the analysis.
While the posteriors of $K$ peak in all models very close to the GR value, the presence of $K$ increases the $68\,\%$ confidence interval of $M$ and $\varepsilon$ by roughly a factor $5$.
\par 
In models with RZ parameters beyond $\varepsilon$, one still finds that the posteriors of all parameters have their maximum very close to the GR values, but their shapes can be very complex.
The RZ posteriors of model$_2$ (Fig.~\ref{fig_modelMea0b0}) are very steep around the GR values, and show little support further away.
In contrast, those of model$_3$, shown in Fig.~\ref{fig_modelMea1b1}, are clearly less constraining for $a_1$ and $b_1$ and admit a small secondary maximum.
A look at the contour plots reveals the strong correlations between certain RZ parameter combinations that produce QNMs very similar to those of Schwarzschild.
\par 
The most complex posteriors are those of model$_{K2}$ in Fig.~\ref{fig_modelMea0b0K_1}.
As can be seen, the posteriors include the GR values, but also admit very small secondary maxima and even stronger correlations between the parameters.
\subsubsection{QNMs and Accuracy}\label{discussion-qnm-accuracy}
Since most of the models studied here have more than two free parameters, it is reasonable to quantify the benefit of measuring multiple QNMs. 
Including the $l=3$ fundamental mode and its first overtone ($n=0, 1$) in addition to the $l=2$ fundamental mode and its first overtone improves the parameter estimation, though the individual improvements vary with the model and the assumed QNM measurement errors.
Since this can be seen clearly for the reconstructed potentials, we discuss this aspect for some models in Sec.~\ref{discussion-potentials} in more detail.
\par 
Decreasing QNM measurement errors by an order of magnitude  provides the strongest improvements.  
QNMs measured with $1\,\%$ errors allow for constraining the posteriors within the prior bounds for all models. 
For model$_{2K}$, which has five free parameters, the $l=3$ QNMs have to be used as well to achieve this.
For some of the other higher dimensional models, we also find non-trivial secondary maxima for the posteriors when the less precise QNMs are used.
\subsubsection{Scaling with Relative Errors}
When the posteriors are well constrained within the limits of the priors and are not multi-modal, we also look at whether our results scale with the relative errors with which we assume that the QNMs are measured.
To this purpose, we relate the width of the $68\,\%$ credible interval of the reconstructed parameters $P_i$, which we denote by $2 \times \sigma_{P_i}$, with the constant relative errors that we have assumed for a given set of QNMs, which we denote by $\delta_\text{QNM}$.
Although there are some minor variations between different models, we generally find
\begin{align}
\frac{\sigma_{P_i}}{\delta_\text{QNM}} \approx \text{constant}_i.
\end{align}
We have also verified this scaling for relative QNM errors of $0.1\,\%$.
The scaling is valid for model$_1$ and model$_{K1}$, while the higher dimensional models are more prone to presenting secondary peaks when the QNM precision is of $\pm 10\,\%$, in which case the notion of a credible interval becomes less clear/relevant.
For the same reasons, we note that the scaling is instead not expected to hold in the other extreme, i.e. for large (i.e.  $100\,\%$) relative errors.
\subsubsection{Priors}\label{discussion-priors}
As expected, we find that the posteriors are more constrained in the lower dimensional models than in the higher dimensional ones, if one assumes the same spectrum
as data.
Since  $M$ is the leading order parameter of the RZ metric, it is generally the best constrained one,
and the width of its posterior distribution is typically smaller than the already tight prior that
we assume on it. This is particularly evident for  simple models, e.g. for model$_1$ in Fig.~\ref{fig_modelMe},
where $M$ has an $68\,\%$ credible interval of $[0.995, 1.005]$.
\par 
Regarding the RZ parameters and $K$, 
for which we recall that we assume large flat priors (respectively $[-1,1]$ and $[-2,8]$),
the posteriors are generally constrained to be well within the priors, i.e. our results are robust. Only for model$_{K2}$ one finds 
tails that tend to extend
 outside the prior ranges for $a_0$ and $b_0$. However,
as we have already mentioned, $a_0,\,b_0\gtrsim 1$ are 
very likely disfavored by GW observations of the inspiral and X-ray tests~\cite{Cardenas_Avendano_2020}, and possibly by other observables not directly related to QNMs (e.g. gravitational redshift, geodesic motion, etc.).
For all of the  cases shown here, the RZ posteriors peak around their Schwarzschild values, and also the posteriors of $K$ peak at the expected GR value $K=3$.
What RZ parameters  can be best constrained depends however, to some extent, on the assumed set of measured QNMs and their errors.
For instance, for some of the higher dimensional models the posterior bounds are less stringent if the errors of the measured QNMs are
10\,\%,  especially if the $l=3$ QNMs are not measured. In that case, at least for some parameters, the posterior widths may even be comparable with the
priors. In other cases, e.g. for model$_{K2}$ shown in Fig.~\ref{fig_modelMea0b0K_1}, the posteriors
have secondary maxima and present strong non-trivial correlations between the RZ parameters (even though it is unclear if the secondary maxima
correspond to RZ metrics describing non-pathological black holes).
\subsection{Reconstruction of Potentials and Metric}\label{discussion-potential-metric}
In the following, we first discuss the reconstruction of the potentials in Sec.~\ref{discussion-potentials}, before addressing the reconstruction of the metric functions in Sec.~\ref{discussion-metrics}.
The results of an additional model in which we enforce the PPN constraints is discussed separately in Sec.~\ref{discussion-ppn}.
\subsubsection{Effective Potentials}\label{discussion-potentials}
The reconstructed effective potentials $V_2(r)$ and $V_3(r)$ are shown in the right top panels of Figs.~\ref{fig_modelMe}, \ref{fig_modelMea0b0}, \ref{fig_modelMea1b1} and \ref{fig_modelMeK}, as well as in Fig.~\ref{fig_modelMea0b0K_2}.
The quality of the reconstruction is clearly related to how well the RZ parameters can be determined.
Since this depends in turn on the underlying model being used, it is not surprising that the potential obtained from model$_1$, shown in Fig.~\ref{fig_modelMe}, is more precisely reconstructed than the one for model$_{K2}$, shown in Fig.~\ref{fig_modelMeK}. 
Since the QNM measurement errors play a major part in how well the parameters can be recovered, we note that the higher dimensional models can have
reconstructed potentials as good as lower dimensional models, if the latter use less precise QNMs. 
\par
As expected from the asymptotic behavior of the RZ metric, the uncertainty in the potentials is minimal for large $r$, because in that region the behavior is dominated by $M$ only.
Since the QNMs are related to the potential at its peak, it is not surprising that the potential becomes drastically less determined away from the maximum, when one approaches the horizon. This is especially the case for higher dimensional models.
\par 
Because adding the $l=3$ QNMs improves the reconstruction of the parameters, one might naively expect a difference between the $l=2$ and $l=3$ potentials according to whether the $l=3$ QNMs have been used or not. 
However, because both potentials depend on the same number of parameters and are constructed almost identically, the impact of including the $l=3$ QNMs depends on the specific model. 
When comparing the reconstructed potentials of model$_1$ in Fig.~\ref{fig_modelMe} with those of model$_{K1}$ in Fig.~\ref{fig_modelMeK}, one sees that the $l=2$ QNMs recover the $l=2$ and $l=3$ potentials with comparable precision for the first model, but the $K$ dependency in model$_{K2}$ makes the $l=3$ potential less constrained. 
However, when adding the $l=3$ QNMs, we find that the reconstruction becomes comparable also for model$_{K2}$. 
\par
We also  note that because the reconstructed potentials present a single maxixum, using the WKB method is indeed justified.
\subsubsection{Metric Functions}\label{discussion-metrics}
The reconstructed metric functions $g_{tt}(r)$ and $g_{rr}(r)$ are shown in the bottom right panels in Figs.~\ref{fig_modelMe}, \ref{fig_modelMea0b0}, \ref{fig_modelMea1b1} and \ref{fig_modelMeK}, as well as in Fig.~\ref{fig_modelMea0b0K_2}. 
The relatively small uncertainties for large values of $r$ are expected by construction, because the RZ metric approaches the Schwarzschild metric asymptotically and $M$ is well constrained. 
Because the information obtained from the QNMs originates from the region around the maximum of the potential, the metric is also well reconstructed there.
As  for the potentials,  the reconstruction of the metric functions also shows some non-trivial differences throughout the different models and QNM subsets.
For models that have $\varepsilon$ as the only RZ parameter,  the reconstruction is similar, but there are significant differences when one includes $b_0$ or $b_1$.
This can be seen most drastically when comparing the results shown in Fig.~\ref{fig_modelMe} with the ones in Fig.~\ref{fig_modelMea1b1}. 
This finding can be explained with a closer look at the structure of $g_{tt}(r)$ and $g_{rr}(r)$ provided in Sec.~\ref{RZ-metric}, which reveals that $g_{tt}(r)$ only depends on $\varepsilon$ and $a_0$ or $a_1$, while $b_0$ or $b_1$ only appear in $g_{rr}(r)$.
The additional degree of freedom of $g_{rr}(r)$ causes its less precise reconstruction.
\subsection{PPN Constraints}\label{discussion-ppn}
The RZ parameters used in model$_3$ are inspired by the PPN constraints $|a_0|, |b_0| \sim 10^{-4}$, which would allow one to set those parameters essentially to zero.
While these bounds may not hold for all alternative theories of gravity,
as discussed in Sec.~\ref{RZ-metric}, we consider them here for comparison
with previous work assuming them \cite{paper8,konoplya2020general,Cardenas_Avendano_2020}. 
Since $a_0=b_0=0$, model$_3$ includes the higher order parameters $a_1$ and $b_1$  with flat priors between $[-1,1]$.
The parameter estimation for this model, shown in Fig.~\ref{fig_modelMea1b1}, is more challenging than for model$_2$, which is shown in Fig.~\ref{fig_modelMea0b0}. 
For this reason, we only report results for the optimistic case of spectrum$_2$ (i.e. with small relative errors of $ 1\,\%$ on the $l=2$ and $l=3$ modes). 
Indeed, for less precise (i.e. 10\%) QNMs or with $l=2$ modes only, we could not constrain all parameters completely within the priors. 
This may occur because $a_1$ and $b_1$ appear as higher order parameters, hence deviations of the potentials and metric functions
away from the Schwarzschild baseline only grow significantly close to the horizon. 
Overall, our results show that even the higher order RZ parameters can be constrained by using QNMs, but only under optimistic conditions (i.e., multiple and precise QNM measurements). 
\subsection{Alternative Inverse QNM Approaches}\label{discussion-alt-inv}
Finally, we also comment briefly on the differences between the present work and other recent related efforts on the inverse QNM spectrum problem of compact non-rotating objects \cite{paper2,paper5,Konoplya:2018ala,paper6,paper8}. 
WKB theory comes in many realizations, and has been applied in different ways depending on the underlying type of QNM spectrum. 
In the case of ultra-compact horizonless objects, for a review on which we refer to \cite{Cardoso:2017cqb}, one finds that there exist long lived trapped modes \cite{1991RSPSA.434..449C,1994MNRAS.268.1015K}. 
For these systems, it is possible to use a generalized Bohr-Sommerfeld rule to describe the spectrum \cite{2014PhRvD..90d4069C,paper1}, and furthermore to invert it in order to constrain the potential \cite{paper2,paper5}. 
While the potentials and QNMs are qualitatively different from Schwarzschild for those objects, the method itself does not require a metric or any type of arbitrary parametrization of the potential. 
However, as a trade-off, the typical number of QNMs required for this method to work is large, and the reconstruction in general not unique; indeed, both features
are typical hallmarks of inverse spectrum problems. 
The advantage of this approach is nevertheless that the reconstructed potentials could have details that are not described by a finite set of metric parameters, which is an intrinsic limitation when following a parametrization approach for inverse problems.  
By following a related approach, it has also been possible to constrain potentials from Hawking radiation \cite{paper7}. 
In Ref.~\cite{Konoplya:2018ala}, the higher order WKB method has been combined with a Morris-Thorne ansatz for the metric to approximate wormholes by using their high frequency QNM spectrum as assumed data. 
\subsection{Possible Extensions}\label{discussion-extensions}
We consider the  work presented here  as a proof of principle effort, which quantifies how well QNMs can be used to constrain black hole metrics that deviate from GR. 
A treatment of the full problem beyond our  toy model requires the knowledge of the field equations of theories beyond GR, which are obviously unavailable in a theory agnostic approach such as ours. 
Another limitation is our focus on non-rotating black holes, since binary black hole mergers will always produce a spinning remnant~\cite{spin1,spin2,spin3}. 
While the RZ metric has also been generalized to describe rotating black holes in Ref.~\cite{Konoplya:2016jvv}, the lack of theory agnostic field equations in the rotating case makes it less clear how to proceed in this direction. 
One possibility would be to work in terms of a slow rotation approximation, which for the axial sector has been studied in Ref.~\cite{2015EPJC...75..560P}. 
Another possible extension of this work may be a Bayesian comparison between different realizations of the RZ metric (i.e. RZ metrics with
different numbers of parameters), to determine the optimal number of free parameters needed to describe a given set of QNM data.
Also, it may be beneficial to incorporate analytic constraints similar to those discussed in Sec.~\ref{param_space} on the RZ parameters directly on the MCMC sampling (using rejection methods). This would allow for a larger prior parameter space, which would be of interest in situations where the data 
are not very informative and when even more RZ parameters are considered.

Finally, an extension of the present framework to incorporate other black hole constraints, like the size of the shadow as observed by the Event Horizon Telescope (EHT) collaboration, is currently in preparation~\cite{shadow_paper}.

\section{Conclusions}\label{conclusions}
Connecting the rising field of experimental gravitational wave physics with fundamental theoretical problems is among the most promising research avenues
in gravitational physics.
In this work we have demonstrated, as a proof of principle, how well the observation of black hole QNMs by gravitational interferometers can be used to constrain the spacetimes of non-rotating black holes. 
By studying several realizations of the RZ metric, as well as an additional degree of freedom of the effective perturbation potential, we have explicitly connected QNMs with phenomenological parameters characterizing deviations from GR. 
Since real experiments cannot observe the full QNM spectrum with infinite precision, we have limited our study to the the $l=2$ and $l=3$ modes, 
considering  both the fundamental mode and the first overtone ($n=0$ and $n=1$), 
and we have assumed several possible measurement errors (between 1\,\% and 10\,\%) for the QNM frequencies and decay times, to mimic the 
effect of various gravitational wave detectors.
\par 
With this setup,  knowledge of the $l=2$ fundamental and first overtone modes is already enough to constrain models with two or three free parameters. 
The more involved models including up to five free parameters require also the corresponding $l=3$ modes for a reasonable parameter estimation. 
As expected, the largest improvement in the parameter estimation is achieved when the QNMs are known with higher precision. 
In this situation, in spite of the the limited number of QNMs, it is possible to constrain even the higher dimensional parametrization models. 
Besides the reconstruction of the metric parameters, we have also  quantified and visualized the errors on the corresponding potentials and metric functions. 

While the general problem of rotating black holes is conceptually and computationally far from trivial, we have demonstrated here that Bayesian parameter estimation and the higher order WKB method provide a suitable framework at least for the non-rotating limit. 
Overall our results suggest that studying the inverse QNM problem is very promising even in the presence of finite number of QNM measurements, and allows for using the ringdown to put constraints on parametrized black holes in gravitational theories beyond GR. 
Since other observational approaches, e.g. shadows obtained by the EHT or X-ray spectroscopy, may also put constraints on the same parametrized black hole geometries, it would be interesting to combine them with QNM bounds. We will address this problem more thoroughly in future work.

\begin{acknowledgments}
We thank Luciano Rezzolla and Prashant Kocherlakota for useful discussions on the RZ parameter space and valuable feedback on the manuscript. 
Furthermore we also thank Kostas D. Kokkotas for sharing his insights on several aspects of our work. 
We also want to thank the anonymous referee for their valuable comments, which have strengthened this work considerably.  
We acknowledge financial support provided under the European Union's H2020 ERC Consolidator Grant
``GRavity from Astrophysical to Microscopic Scales'' grant agreement no. GRAMS-815673. 
\end{acknowledgments}

\bibliography{literature}

\end{document}